\documentclass[sigconf]{acmart}
\AtBeginDocument{%
  }

\setcopyright{acmlicensed}
\copyrightyear{2018}
\acmYear{2018}
\acmDOI{XXXXXXX.XXXXXXX}
\acmConference[Conference acronym 'XX]{Make sure to enter the correct
  conference title from your rights confirmation email}{June 03--05,
  2018}{Woodstock, NY}
\acmISBN{978-1-4503-XXXX-X/18/06}

\usepackage{subcaption}
\usepackage{xcolor}
\usepackage{tcolorbox}
\usepackage{listings}
\usepackage{graphicx}
\usepackage{csquotes}
\usepackage{refcount}

\usepackage{multirow}
\usepackage{booktabs}

\begin{document}

\title{AutoCas: Autoregressive Cascade Predictor in Social Networks via Large Language Models}


\author{Yuhao Zheng}
\affiliation{%
  \institution{University of Science and Technology of China}
  \city{Hefei}
  \country{China}}
\email{yuhaozheng@mail.ustc.edu.cn}

\author{Chenghua Gong}
\affiliation{%
  \institution{East China Normal University}
  \city{Shanghai}
  \country{China}}
\email{chenghuagong@stu.ecnu.edu.cn}
\authornote{Co-frst authors with equal contribution.}

\author{Rui Sun}
\affiliation{%
  \institution{Hefei University of Technology}
  \city{Hefei}
  \country{China}}
\email{2021217844@mail.hfut.edu.cn}

\author{Juyuan Zhang}
\affiliation{%
  \institution{University of Science and Technology of China}
  \city{Hefei}
  \country{China}}
\email{zhangjuyuan2020@mail.ustc.edu.cn}

\author{Liming Pan}
\affiliation{%
  \institution{University of Science and Technology of China}
  \city{Hefei}
  \country{China}}
\email{pan_liming@ustc.edu.cn}

\author{Linyuan L\"u}
\affiliation{%
  \institution{University of Science and Technology of Chinay}
  \city{Hefei}
  \country{China}}
\email{linyuan.lv@ustc.edu.cn}
\authornote{Corresponding authors}

\renewcommand{\shortauthors}{Yuhao Zheng et al.}


\begin{abstract}
    Popularity prediction in information cascades plays a crucial role in social computing, with broad applications in viral marketing, misinformation control, and content recommendation. However, information propagation mechanisms, user behavior, and temporal activity patterns exhibit significant diversity, necessitating a foundational model capable of adapting to such variations. At the same time, the amount of available cascade data remains relatively limited compared to the vast datasets used for training large language models (LLMs).
    Recent studies have demonstrated the feasibility of leveraging LLMs for time-series prediction by exploiting commonalities across different time-series domains. Building on this insight, we introduce the Autoregressive Information Cascade Predictor (AutoCas), an LLM-enhanced model designed specifically for cascade popularity prediction. Unlike natural language sequences, cascade data is characterized by complex local topologies, diffusion contexts, and evolving dynamics, requiring specialized adaptations for effective LLM integration.
    To address these challenges, we first tokenize cascade data to align it with sequence modeling principles. Next, we reformulate cascade diffusion as an autoregressive modeling task to fully harness the architectural strengths of LLMs. Beyond conventional approaches, we further introduce prompt learning to enhance the synergy between LLMs and cascade prediction. 
    Extensive experiments demonstrate that AutoCas significantly outperforms baseline models in cascade popularity prediction while exhibiting scaling behavior inherited from LLMs.
    Code is available at this repository:~\url{https://anonymous.4open.science/r/AutoCas-85C6}

\end{abstract}






\maketitle
    
\section{Introduction}
    As social media platforms such as Twitter, TikTok, Rednote, WeiBo, and Reddit become increasingly embedded in our daily lives, they enable a continuous creation and widespread dissemination of vast amounts of user-generated content.  User activity patterns within these networks play a pivotal role in shaping the spread of information, often giving rise to large-scale information cascades~\cite{cheng2014can}.
    Effectively modeling these cascades to predict the trajectory of information diffusion is of paramount importance across various real-world applications, including fake news detection~\cite{zhang2023deep}, viral marketing~\cite{leskovec2007dynamics}, epidemic prevention~\cite{zhao2020online}, and recommender systems~\cite{wu2019dual}, among others.
    
    \begin{figure}[htpb]
    \includegraphics[width=0.99\linewidth]{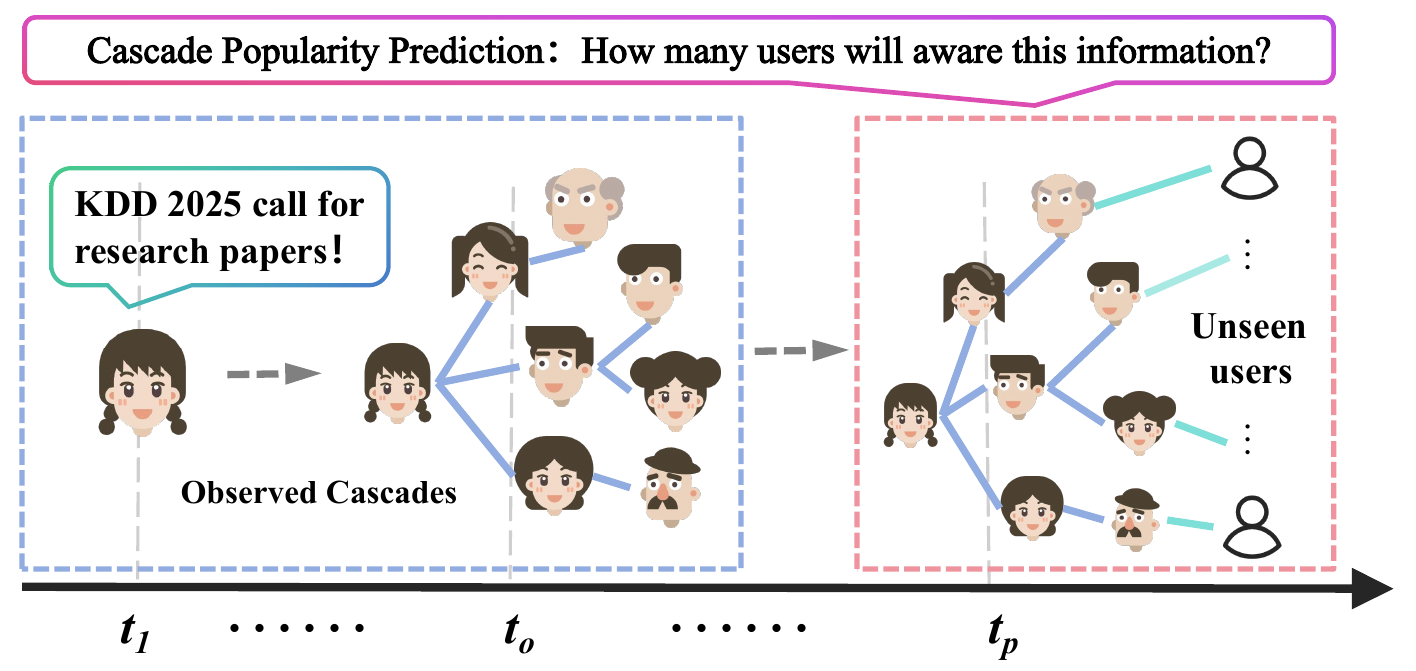}
    \caption{Illustration of cascade popularity prediction.
    Given an information item, the target is to model its diffusion process and predict how many users will aware it.}
    \label{cas}
    \end{figure}
    \begin{figure*}
    \includegraphics[width=0.99\linewidth]{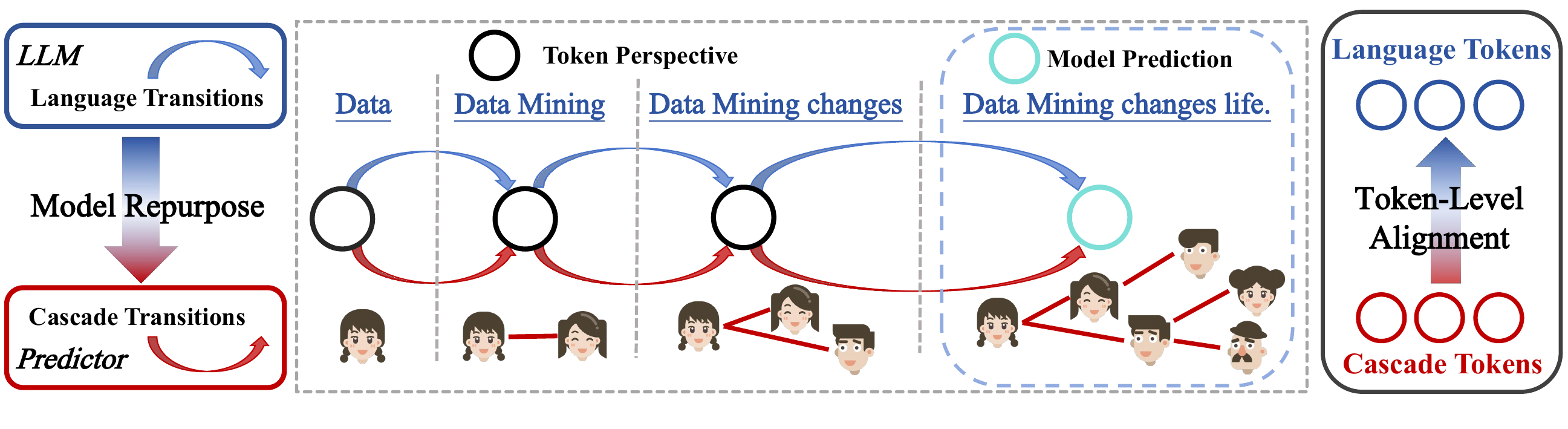}
    \caption{The motivation of adapting large language models for cascade modeling, which demonstrates the process of text generation and cascade diffusion, indicating that both can be viewed as sequence modeling from the token perspective.}
    \label{motivation}
    \end{figure*}
    
    Among information cascade modeling tasks, cascade popularity prediction has attracted widespread attention from both academia and industry~\cite{li2017deepcas,casflow,CasCN,zhou2021survey}.
    An illustration of the information cascade modeling task is shown in Figure~\ref{cas}: given the observed propagation status of an information item, the target is to model its diffusion process and predict the future cascade size~\cite{zhou2021survey}. 
    Previous approaches can be summarized into the following three categories. 
    (1)~\textbf{Feature-based methods}. 
    Early studies~\cite{szabo2010predicting} focus on extracting hand-crafted features from cascades, e.g., structural features, temporal features, user-related features, etc for cascade prediction. 
    (2)~\textbf{Statistics-based methods}.  
    These studies model the information diffusion process via the intensity function of the arrival for incoming messages, e.g., Poisson~\cite{iwata2013discovering} and Hawkes~\cite{rizoiu2018sir} processes.
    (3)~\textbf{Deep learning methods.}
    This type studies explore the potential of deep learning techniques in cascade popularity prediction.
    Most of them integrate the temporal information, structural information, and complex evolution patterns of cascades with RNNs~\cite{casflow}, GNNs~\cite{CasCN,cao2020popularity}, Transformer~\cite{Castformer}, and Neural ODEs~\cite{CasDO}, among others.
    Due to powerful generalization capabilities, deep learning methods have become the current mainstream approach in information cascade modeling.
    
    The success of LLMs highlights the transformative potential of foundational models across diverse fields, including graph~\cite{gong2024survey,liu2023towards}, geography~\cite{roberts2023gpt4geo}, and seismology~\cite{liu2024seislm}. 
    In the context of information cascades, propagation mechanisms, user behavior, and temporal activity patterns exhibit both universal patterns and significant variability, necessitating a foundational model capable of capturing shared patterns while adapting to diverse dynamics. However, the development of such a model is constrained by the limited availability of cascade data, posing challenges for effective training. Fortunately, recent studies in vision~\cite{zhang2024vision}, and time-series~\cite{liu2024autotimes} have demonstrated the feasibility of task reformulation and repurpose LLMs for general task adaptation.
    Therefore, a natural question arises: \textit{Can we adopt powerful LLMs to perform information cascade modeling?}
    
    Intuitively, language and cascade generation share basic commonalities, presenting opportunities to repurpose the off-the-shelf LLMs for cascade modeling.
    As can be seen in Figure~\ref{motivation}, both language generation and cascade diffusion can be regarded as sequence modeling at the token level.
    Similar observations can be found in~\cite{liu2024autotimes}, where time-series and language are unified via token transitions in sequence modeling to repurpose LLMs as forcasters. 
    Ideally, if cascade data can be tokenized into cascade tokens and aligned with the linguistic embedding space, it would be feasible to repurpose LLMs as the cascade predictor. 
    Nevertheless, unlike text sequence and time-series patch, the cascade data possesses the inherent local topological structure, complex global interactions, and the dynamics of cascade diffusion.
    Therefore, one challenge of applying LLMs to information cascade modeling lies in~\textit{whether cascade data can be tokenized to make it compatible with LLM architecture?}
    
    Prior studies have revealed that powerful generalization of LLMs largely stems from the autoregressive modeling paradigm~\cite{wang2022language}.
    In essence, autoregressive modeling involves predicting the next value in a sequence based on previous data points~\cite{gregor2014deep}, which is highly compatible with language generation and has already been the best practice for building LLMs currently~\cite{achiam2023gpt,bai2023qwen}.
    To maximizing its potential, studies which repurpose LLM as the foundational base in fields such as vison~\cite{peng2024multi,rajasegaran2025empirical} and time-series~\cite{liu2024autotimes} endeavor to reconfigure task formulations, transforming them into autoregressive templates.
    Notably, the expansion of information cascade is dependent on its historical states, making autoregressive cascade modeling feasible.
    Consequently, the key to effectively utilizing LLMs for cascade modeling hinges on~\textit{whether the cascade diffusion process can be reformulated as autoregressive modeling?}

    To further investigate the diffusion trends of information cascade, we place greater emphasis on popularity prediction based on cascade modeling.
    Compared to conventional deep learning methods, the repurpose of LLM as predictor is likely to cause inadequate task adaptation for cascade popularity forecast.
    Inspired by the remarkable success in NLP, prompting learning~\cite{zhang2023prompt} has advanced in facilitating the integration of LLMs into specific tasks across various domains~\cite{sun2023graph,gong2024self}.
    Since LLMs exhibit preferences towards textual prompts, appropriate textual prompt templates should be deeply coordinated with the cascade data for subsequent predictions. 
    We expect that cascade data can be prompted by relevant contexts, so this last obstacle lies in
    ~\textit{whether we can integrate LLMs into cascade popularity prediction task via prompt learning?}
    
    In this paper, we repurpose the LLMs as cascade predictors, and present a novel framework called AutoCas.
    Technically, we first tokenize the information cascade data into a sequence of cascade tokens which preserves local structural patterns and global complex interactions.
    To align with LLM architecture, we project the cascade tokens into the embedding space of language, reformulate the cascade diffusion as an autoregressive process, thereby fully leveraging the modeling capabilities of LLM. 
    Thanks to autoregressive modeling, AutoCas can flexibly accommodate any new observation time without the need for retraining, unlike existing methods that require retraining for each distinct observation time.
    Going beyond the traditional cascade prediction paradigm, we creatively propose the cascade prompt learning via designing textual prompt to facilitate the deeper integration of LLMs into cascade popularity prediction.
    Our contributions can be mainly summarized as follows:
    \begin{itemize}
    \item 
    We view inforamtion cascade diffusion from a fresh perspective, aligning it with language generation at the token level, and introduce AutoCas, a novel approach to repurpose the LLMs as cascade predictors.
    To our best knowledge, AutoCas is the first LLM-based framework in this domain.
    \item 
    We reformulate cascade modeling in autoregressive paradigm to unleash the potential of LLM architecture, and introduce cascade prompt learning, which advances the traditional paradigm via textual prompts to promote cascade popularity prediction.

    \item 
    We conduct extensive experiments on large-scale real-world datasets to evaluate AutoCas. The results show that AutoCas significantly outperforms other state-of-the-art competitors in cascade popularity prediction task and exhibits scaling behavior empowered by LLMs.
    \end{itemize}

\section{Related Works}
    
\subsection{Information Cascade Modeling}
    Information cascade modeling can be primarily categorized into two types: micro-level~\cite{qiu2018deepinf,yang2019neural} and macro-level~\cite{li2017deepcas,CasCN}.
    The former focuses on predicting next affected user, while the latter concentrates on the overall trends, such as its popularity or outbreak status.
    In this paper, we focus on the cascade popularity prediction and categorize existing methods into three types:
    
    \noindent \textbf{Feature-based methods.}
    These works focus on making hand-crafted features for cascade and conducting popularity prediction via traditional machine learning approaches~\cite{szabo2010predicting}.
    However, feature-based methods heavily rely on the expert knowledge, has high customization costs, and exhibits limited generalization and suboptimal performance~\cite{qiu2018deepinf}. 

    \noindent \textbf{Statistics-based methods.}
    These studies assume that information diffusion follow a specific probability statistical model, such as the Poisson process~\cite{iwata2013discovering}, Hawkes process~\cite{rizoiu2018sir}. 
    Statistics-based methods are interpretable but have strong parametric assumptions, making them unsuitable for real-word applications~\cite{zhou2021survey}.
    
    \noindent \textbf{Deep leaning methods.}
    These methods adopt deep learning techniques to promote cascade popularity prediction.
    Early representative works, such as DeepCas~\cite{li2017deepcas} and DeepHawkes~\cite{cao2017deephawkes}, focus on capturing the temporal dynamics of cascades via RNN or LSTM.
    Considering the topology in information cascade, GNNs have been introduced to capture local structural patterns. 
    For example, CasCN~\cite{CasCN} and CoupledGNN~\cite{cao2020popularity} adopt variant GNNs to model the interactions between users and the spread influence.
    Apart from local structure, CasFlow~\cite{casflow} introduces the social network as global context to enhance popularity prediction.
    Advanced techniques, such as VAEs~\cite{zhou2020variational,wang2021dydiff}, Transformers~\cite{Castformer,yu2022transformer}, and Neural ODEs~\cite{rubanova2019latent,cheng2024information}, have been further explored in cascade modeling.
    For more comprehensive reviews, please refer to~\cite{zhou2021survey,li2021capturing}.

    \noindent \textbf{Towards LLM-based methods.}
    Due to the strong generalization capabilities, various fields such as vision~\cite{awais2025foundation,zhang2024vision} and time-series~\cite{liu2024autotimes,liang2024foundation} are renovates by the general frameworks based on LLMs. 
    To our best knowledge, this is the first attempt to introduce an LLM-based method in information cascade modeling.

\subsection{Autoregressive Modeling}
    Autoregression is a fundamental concept in sequence modeling, which uses observations from previous time steps to predict the next value~\cite{gregor2014deep}.
    This paradigm, which provides fine-grained supervision, has become the best practice for training LLMs ~\cite{radford2019language,achiam2023gpt,bai2023qwen} and has also inspired other fields~\cite{van2016pixel,tian2024visual,rajasegaran2025empirical}.
    Here, we briefly categorize existing works into three types:
    
    \noindent \textbf{RNN-based methods.}
    Early studies perform the autoregressive modeling based on RNN variants, and achieve success across various domains~\cite{ranzato2014video,van2016conditional,van2016pixel}.
    However, these approaches come with imitations of RNNs~\cite{lipton2015critical}, including low computational efficiency and limited capability in long-distance dependencies.
    
    \noindent \textbf{Transformer-based methods.}
    Following the introduction of Transformer~\cite{vaswani2017attention}, the potential of autoregressive modeling has been further explored, with representative works including iGPT~\cite{chen2020generative}, Autoformer~\cite{wu2021autoformer} and VAR~\cite{tian2024visual}.
    Currently, transform-based methods dominate the field of autoregressive modeling.
    
    \noindent \textbf{LLM-based methods.}
    Built upon Transformer architecture, LLMs with large-scale parameters are pretrained on massive datasets, demonstrating superior capabilities in autoregressive modeling.
    Therefore, researchers attempt to investigate the feasibility of reusing LLMs for autoregressive modeling.
    For example, Toto~\cite{rajasegaran2025empirical} treats videos as sequences of visual tokens and reuses the LLMs as backbones to autoregressively predict future tokens.
    Similar ideas have also been applied in the field of time-series forecasting~\cite{liu2024autotimes}.

\subsection{Prompt Learning}
    Prompt learning~\cite{liu2023pre} has emerged as a novel learning paradigm to adapt LLMs to specific tasks by designing textual prompts.
    Due to the widespread application of LLMs, designing sophisticated textual prompts for specific tasks has become the research hotspot across various fields~\cite{gong2024self}.
    For example, 
    LLM4NG~\cite{yu2023empower} designs the class-level semantic prompt templates based on text-attribute on graphs to facilitate the node classification in few-shot scenarios.
    Autotimes~\cite{liu2024autotimes}  introduces textual timestamps of time-series to enhance LLM-based forecasting.
    Building on advanced prompting techniques, deft textual prompts ~\cite{jin2023time,liu2024unitime} for time-series are further explored.
    Since LLMs have not yet fully entered the field of information cascade modeling, studies on prompt learning for cascade data are still lacking. 
    However, textual information is prevalent in cascade diffusion process, so designing cascade prompt templates based on textual information for cascade popularity prediction holds significant promise.

\section{Preliminaries}
    \noindent \textbf{Information cascade graph.}
    Given an information item and all its diffusion, the information cascade graph is defined as $\mathcal{G}_{c}(t)=\{ \mathcal{V}_{c}(t),\mathcal{E}_{c}(t)\}$, where $\mathcal{V}_{c}(t)$ is the set of users participated in the cascade diffusion up to time $t$, and $\mathcal{E}_{c}(t)$ is the set of edges between users, each edge $e_{i,j} = (v_i,v_j,t_k)$ denotes that user $v_i$ retweets the information to user $v_j$ at time $t_k$.

    \noindent \textbf{Global context graph.}
    The global context graph $\mathcal{G}_{g}=\{\mathcal{V}_{g},\mathcal{E}_{g}\}$ represents the context of cascade scene, e.g., social networks or citation networks, where $\mathcal{V}_{g}$ represents the set of all users and $\mathcal{E}_{g}$ denotes the relationships between them in the context.

    \noindent \textbf{Cascade popularity prediction.}
    Given the information cascade graph $\mathcal{G}_c$ and global context graph $\mathcal{G}_g$, our target is to predict the future cascade popularity $\mathcal{P}(t_p) = |\mathcal{V}_c(t_p)|$ at a prediction time $t_p$ based on the observation time $t_o$ ($t_o\ll t_p$).

    \begin{figure*}[htbp]
    \centering
    \includegraphics[width=0.98\linewidth]{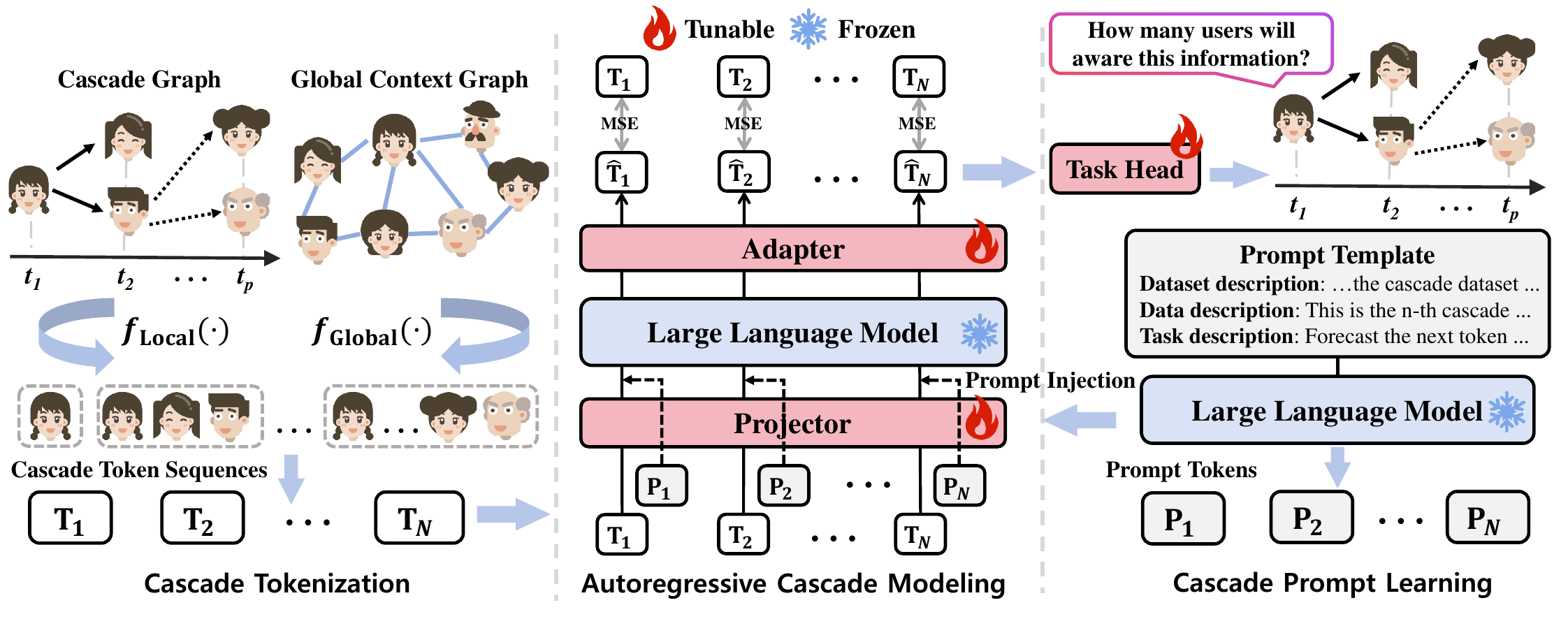}
    \caption{The overview framework of AutoCas. (1) Local and global embeddings extraction and cascade token sequences construction;
    (2) Autoregressive cascade modeling with pre-trained large language model as backbone;
    (3) Textual prompt template design and injection.}
    \label{framework}
    \end{figure*}
    
\section{Methodology}
    This section introduces details of our proposed AutoCas.
    AutoCas consists of three modules: (1) cascade tokenization, (2) autoregressive cascade modeling, and (3) cascade prompt learning.
    The overall framework is shown in is shown in Figure~\ref{framework}.
    
    \subsection{Cascade Tokenization}
    The key to adapting LLMs to information cascade modeling lies in whether cascade data can be tokenized.
    Unlike nature language, the complex patterns of cascade data pose challenges to tokenization in three aspects:
    (1) The local topological pattern of the cascade, 
    (2) The global position of cascade within its context,
    (3) The dynamic of cascade evolution over time.
    Therefore, cascade tokenization should involve the aforementioned three type information simultaneously.
    In this paper, we introduce local embeddings and global embeddings to capture the local topology and global context of cascade data, and construct cascade token sequences to capture the evolutionary dynamics.
    
    \noindent \textbf{Local embeddings.}
    For local structure, the method of extracting topological patterns of cascade is flexible. 
    Traditional methods like DeepWalk~\cite{perozzi2014deepwalk}, LINE~\cite{tang2015line}, and Node2Vec~\cite{grover2016node2vec} are feasible, and furthermore, various advanced GNN-based approaches~\cite{song2022graph} are also available.
    Given the cascade graph $\mathcal{G}_c=\{\mathcal{V}_c,\mathcal{E}_c\}$, the local embedding function $f_{\mathrm{local}}(\cdot)$ is to embed each user according to its topological pattern within the cascade diffusion:
    \begin{equation}
        \mathbf{LE} = f_{\mathrm{local}} ( \mathcal{G}_{c} ) = f_{\mathrm{local}} ( \mathcal{V}_{c},\mathcal{E}_{c} ).
    \end{equation}
    We follow the previous study~\cite{zhou2024modeling} and implement $f_{\mathrm{local}}(\cdot)$ with spectral graph wavelets~\cite{donnat2018learning}.
    Specifically, we calculate the spectral wavelets based on graph Laplacian with heat kernel function to obtain local embedding $\mathbf{LE}_{i}$ for user $v_i$.  
    
    \noindent \textbf{Global embeddings.}
    The global context graph  encompasses the propagation background of users involved in cascades, from which rich semantic information can be extracted~\cite{casflow}.
    Given the context graph $\mathcal{G}_{g}= \{ \mathcal{V}_{g}, \mathcal{E}_{g} \}$, the global embedding function $f_{\mathrm{global}}(\cdot)$ is to embed each user according to its global position within the context:
    \begin{equation}
        \mathbf{GE} = f_{\mathrm{global}} ( \mathcal{G}_{g} ) = f_{\mathrm{global}} ( \mathcal{V}_{g}, \mathcal{E}_{g}  ).
    \end{equation}
    For large-scale context graphs with millions of nodes, generating position embeddings for each user is challenging due to the high computational overhead.
    To this end, we implement $f_{\mathrm{global}}(\cdot)$ with structural learning embedding method based on sparse matrix factorization~\cite{zhang2019prone} to get global embedding $\mathbf{GE_{i}}$ for user $v_i$, which is proved to be to scalable and efficient in previous studies~\cite{casflow,zhou2024modeling}.
    
    Local and global pattern fusion has been proved to be effective in cascade modeling~\cite{CasDO,casflow,zhou2024modeling}.
    For each user $v_i$, we fuse its local and global embeddings to obtain the final user embedding $\mathbf{H}_{v_i}$:
    \begin{equation}
        \mathbf{H}_{v_i} = \mathtt{Concat} \left( \mathbf{LE}_{v_i},\mathbf{GE}_{v_i} \right)\in\mathbb{R}^{d},
    \end{equation}
    where $\mathtt{Concat}(\cdot)$ denotes the concatenation operation and $d$ is the user embedding dimension.
    
    \noindent \textbf{Cascade token sequence.} 
    To capture the temporal dependence and evolution dynamics, we divide the cascade diffusion process into $N$ time patches.
    Cascade token characterizes the user snapshot of cascade within a time patch. 
    Given the set of users $\mathcal{V}_c(t_n)=\{v_1,v_2,\cdots,v_i\}$ who have participated in the information cascade until the $n-th$ time patch , we construct the corresponding cascade token $\mathbf{T}_n$ via the pooling operation:
    \begin{equation}
        \mathbf{T}_{n} = \mathtt{Pooling}\left(
        \mathbf{H}_{v_i}
        \right), v_i \in \mathcal{V}_c(t_n).
    \end{equation}
    Here, $\mathtt{Pooling(\cdot)}$ can take various forms, ranging from the simple sum and concatenation pooling.
    In our implementation, we concatenate user embeddings according to the propagation order for convenience. 
    Thus, we set a maximum length $l$ to limit the token size and apply default padding, and $\mathbf{T}_{n}\in\mathbb{R}^{l \times d}$ is then flattened into $\mathbf{T}_{n}\in\mathbb{R}^S$.
    After tokenization at each time patch, we can finally obtain the cascade token sequence:
    \begin{equation}
        \mathcal{T} = \left\{\mathbf{T}_{1},\mathbf{T}_{2},\cdots,\mathbf{T}_{N}\right\}\in\mathbb{R}^{N \times S},
    \end{equation}
    which can subsequently be seamlessly integrated into autoregressive modeling based on the LLM architecture.

    \subsection{Autoregressive Cascade Modeling}
    When reusing the LLMs for cascade modeling, the cascade data, although tokenized, still has to be aligned to language tokens.
    Therefore, we adopt $\mathtt{Projector}(\cdot):\mathbb{R}^S\mapsto\mathbb{R}^D$ to map the cascade tokens into the embedding space of language tokens:
    \begin{equation}
        \left\{ {\mathbf{Z}}_{1},
        {\mathbf{Z}}_{2},..., {\mathbf{Z}}_{N}
        \right\}
        =\mathtt{Projector}\left(
        \left\{\mathbf{T}_{1},
          \mathbf{T}_{2},\cdots,
          \mathbf{T}_{N}
        \right\}
        \right),
    \end{equation}
    where $D$ is consistent with the dimension of the LLM.
    
    Based on large-scale autoregressive pre-training, prevalent LLMs can effectively predict the next token based on the preceding tokens.
    To fully unleash the potential of LLM architecture, we redefine the cascade diffusion in a fully consistent approach.
    We feed the into projected cascade tokens into the intermediate layers of LLM:
    \begin{equation}
        \left\{ \widehat{\mathbf{Z}}_{2},
        \widehat{\mathbf{Z}}_{3},..., \widehat{\mathbf{Z}}_{N}
        \right\}
        =\mathtt{LLM}\left(
        \left\{\mathbf{Z}_{1},
          \mathbf{Z}_{2},\cdots,
          \mathbf{Z}_{N-1}
        \right\}
        \right).
    \end{equation}
    Then each predicted patch is mapped back by $\mathtt{Adapter}(\cdot):\mathbb{R}^D\mapsto\mathbb{R}^S$ into the space of cascade tokens for supervision:
    \begin{equation}
        \left\{ \widehat{\mathbf{T}}_{2},
        \widehat{\mathbf{T}}_{3},..., \widehat{\mathbf{T}}_{N}
        \right\}
        =\mathtt{Adapter}\left(
        \left\{ \widehat{\mathbf{Z}}_{2},
        \widehat{\mathbf{Z}}_{3},\cdots, \widehat{\mathbf{Z}}_{N}
        \right\}
        \right).
    \end{equation}
    Both $\mathtt{Projector}(\cdot)$ and $\mathtt{Adapter}(\cdot)$ can be implemented through a linear layer or an MLP (Multi-Layer Perceptron).
    Finally, each predicted token is supervised by the token-wise ground truth:
    \begin{equation}
        \mathcal{L}_{\mathrm{MSE}} = \frac{1}{N-1} \sum_{n=2}^N\left\Vert\mathbf{T}_{n}-\widehat{\mathbf{T}}_{n}\right\Vert_2^2.
    \end{equation}
    Notably, we freeze the parameters of the LLM, only tune the parameters of the lightweight projector and adapter, significantly reducing training costs and enabling quick modeling adaptation.
    
    \subsection{Cascade Prompt Leanring}
    When introducing prompt learning, it is necessary to design appropriate prompt templates to facilitate adaptation to specific tasks.
    Given the LLM backbone, textual prompts are more preferred and have been demonstrated as an enhancement in the prediction task across various fields~\cite{liu2024autotimes}.
    Insipred by that, we adopt the textual timestamps as prompts to provide extra semantic information for cascade token $\mathbf{T}_n$:
    \begin{equation}
        \mathbf{P}_n = \texttt{LLM}
        \left(
        \texttt{Prompt\_Template}
        \left(\mathbf{T}_n
        \right)
        \right),
    \end{equation} 
    where $\texttt{Prompt\_Template}(\cdot)$ customizes the text of cascade timestamps and $\mathbf{P}_n$ is the prompt token embedded by LLM.
    The specific template of the prompts we adopt can be found in Appendix \ref{appendixC}.
    Similar to the functionality of position embedding, the prompt token we designed carry the temporal position of cascade during the diffusion process.
    So we naturally inject it into the space of language embedding after the projector:
    \begin{equation}
        \left\{ \widehat{\mathbf{Z}}_{2},
        \widehat{\mathbf{Z}}_{3},..., \widehat{\mathbf{Z}}_{N}
        \right\}
        =\mathtt{LLM}\left(
        \left\{\mathbf{Z}_{1}+\mathbf{P}_{1},\mathbf{Z}_{2}+\mathbf{P}_{2},\cdots,
        \mathbf{Z}_{N}+\mathbf{P}_{N}
        \right\}
        \right),
    \end{equation}
    After prompt injection, we shift our focus back to popularity prediction.
    Since the cascade token retain the semantic of cascade diffusion, we further adopt it to predict the popularity of information item $m$ at the target time $t_p$:
    \begin{equation}
        \widehat{\mathcal{P}}(t_p) = \mathtt{Task\_Head}(\widehat{\mathbf{T}}_{N}),
    \end{equation}
    where $\mathtt{Task\_Head}(\cdot)$ is a two-layer MLP in our implementation.
    We use the mean square logarithmic error (MSLE) as the objective, and the loss is defined as:
    \begin{equation}
        \mathcal{L} = \frac{1}{M} \sum_{m=0}^{M}
        (\log \mathcal{P}_m(t_p)- \log \widehat{\mathcal{P}}_m(t_p))^2,        
    \end{equation}
    where $M$ is the total number of cascades, $\mathcal{P}_m(t_p)$ is the ground truth, i.e., the number of user who retweets the cascade. $\widehat{\mathcal{P}}_m(t_p)$ is the predicted popularity for information cascade item $m$.

    \begin{table}
    \caption{The statistics information of all datasets.}
    \label{datasets}
    \begin{tabular}{c c c c }
    \toprule
    \textbf{Dataset} &  $\mathtt{Weibo}$ & $\mathtt{Twitter}$  & $\mathtt{APS}$ \\
    \midrule
    \textbf{Cascades} & 119,313 & 88,440 &  207,685  \\
    \textbf{Nodes in}$~\mathcal{G}$ & 6,738,040 & 578,913 & 616,316  \\
    \textbf{Edges in}$~\mathbf{G}$ & 15,311,973 & 7,998,380 & 3,304,400 \\
    \midrule
    \textbf{Total Duration} & 24 hour & 32 days & 20 years\\
    \textbf{Avg. popularity} & 240 & 124 & 51 \\
    \textbf{Observed period} & 0.5/1 hour & 1/2 day(s) & 3/5years \\
    \bottomrule
    \end{tabular}
    \end{table}
    
\section{Experiments}
    In this section, we perform experiments on benchmark datasets to evaluate AutoCas and attempt to answer the following questions:
    
    \noindent \textbf{RQ1:} How effective is AutoCas for cascade popularity prediction?
    
    \noindent \textbf{RQ2:} How key components of AutoCas affect its performance?
    
    \noindent \textbf{RQ3:} Does AutoCas exhibit inherited scaling behavior from LLM?
    
    \noindent \textbf{RQ4:} How does AutoCas improve efficiency over other baselines?
    
    \begin{table*}
    \small
    \caption{The performance of cascade popularity prediction over different observation periods.
    T-test is performed for statistical significance of the results.
    The best results of existing methods are highlighted with \textbf{\textcolor[HTML]{006eb8}{blue}}.
    The best and worst performances for AutoCas are marked with \textbf{\textcolor[HTML]{cf2826}{red}} and \textbf{\textcolor{orange}{orange}}, respectively.
    }
    \label{results}
    \begin{tabular}{c | c c | c c | c c | c c | c c | c c}
    \toprule
    \multirow{2}{*}{\textbf{Model}} & 
    \multicolumn{2}{c}{\textbf{Weibo-0.5 Hour}} &
    \multicolumn{2}{c|}{\textbf{Weibo-1 Hour}} &
    \multicolumn{2}{c}{\textbf{Twitter-1 Day}} &
    \multicolumn{2}{c|}{\textbf{Twitter-2 Days}} &
    \multicolumn{2}{c}{\textbf{APS-3 Years}} &
    \multicolumn{2}{c}{\textbf{APS-5 Years}} \\ \cmidrule(r){2-13}
    & MSLE & MAPE & MSLE & MAPE & MSLE & MAPE & MSLE & MAPE & MSLE & MAPE & MSLE & MAPE \\
    \midrule
    Feat-P & 4.455 & 0.391 & 4.001 & 0.398 & 14.792 & 0.961 & 13.515 & 0.983 & 2.382 & 0.316 & 2.348 & 0.352 \\
    Feat-Linear & 2.959 & 0.258 & 2.640 & 0.271 & 9.326 & 0.520 & 6.758 & 0.459 & 1.852 & 0.272 & 1.728 & 0.291 \\
    Feat-MLP & 2.715 & 0.228 & 2.546 & 0.272 & 7.438 & 0.485 & 6.357 & 0.500 & 1.844 & 0.270 & 1.666 & 0.282 \\
    \midrule
    TimeSeries & 3.119 & 0.277 & 2.693 & 0.268 & 8.214 & 0.547 & 6.023 & 0.445 & 1.867 & 0.271 & 1.735 & 0.291 \\
    DeepHawkes & 2.891 & 0.268 & 2.796 & 0.282 & 7.216 & 0.587 & 5.788 & 0.536 & 1.573 & 0.271 & 1.324 & 0.335 \\
    \midrule
    CasCN & 2.804 & 0.254 & 2.732 & 0.273 & 7.183 & 0.547 & 5.561 & 0.525 & 1.562 & 0.268 & 1.421 & 0.265 \\
    CoupledGNN & 2.845 & 0.255 & 2.766 & 0.281 & 7.216 & 0.562 & 5.578 & 0.525 & 1.673 & 0.275 & 1.395 & 0.259 \\
    FOREST & 2.755 & 0.249 & 2.753 & 0.279 & 7.256 & 0.556 & 5.558 & 0.494 & 1.603 & 0.265 & 1.404 & 0.265 \\
    LatentODE & 2.429 & 0.228 & 2.234 & 0.245 & 7.112 & 0.475 & 5.232 & 0.377 & 1.367 & 0.227 & 1.373 & 0.251 \\
    Dydiff-VAE & 2.416 & 0.220 & 2.227 & 0.253 & 7.012 & 0.463 & 5.205 & 0.372 & 1.372 & 0.232 & 1.362 & 0.247 \\
    CTCP & 2.546 & 0.236 & 2.399 & 0.245 & 6.625 & 0.421 & 4.937 & 0.353 & 1.401 & 0.238 & 1.539 & 0.266 \\
    MINDS & 2.589 & 0.232 & 2.478 & 0.236 & 6.514 & 0.431 & 4.757 & 0.339 & 1.425 & 0.241 & 1.418 & 0.236 \\
    CasFlow & 2.404 & 0.224 & 2.281 & 0.221 & 6.955 & 0.456 & 5.144 & 0.362 & 1.364 & 0.246 & 1.352 & 0.234 \\
    CasTFormer & 2.362 & 0.217 & 2.268 & 0.230 & 6.637 & 0.437 & 4.972 & 0.345 & 1.363 & 0.249 & 1.536 & 0.242 \\
     CasDO & \textbf{\textcolor[HTML]{006eb8}{2.308}} & \textbf{\textcolor[HTML]{006eb8}{0.201}} & \textbf{\textcolor[HTML]{006eb8}{2.178}} & \textbf{\textcolor[HTML]{006eb8}{0.214}} & \textbf{\textcolor[HTML]{006eb8}{6.376}} & \textbf{\textcolor[HTML]{006eb8}{0.376}} & \textbf{\textcolor[HTML]{006eb8}{4.511}} & \textbf{\textcolor[HTML]{006eb8}{0.311}} & \textbf{\textcolor[HTML]{006eb8}{1.093}} & \textbf{\textcolor[HTML]{006eb8}{0.207}} & \textbf{\textcolor[HTML]{006eb8}{1.011}} & \textbf{\textcolor[HTML]{006eb8}{0.229}} \\
    \midrule
    AutoCas (GPT-2) & \textbf{\textcolor{orange}{1.242}} & \textbf{\textcolor{orange}{0.177}} & \textbf{\textcolor{orange}{1.169}} & \textbf{\textcolor{orange}{0.152}} & \textbf{\textcolor{orange}{2.415}} & \textbf{\textcolor{orange}{0.249}} & \textbf{\textcolor{orange}{2.376}} & \textbf{\textcolor{orange}{0.240}} & \textbf{\textcolor{orange}{0.764}} & \textbf{\textcolor{orange}{0.139}} & \textbf{\textcolor{orange}{0.681}} & \textbf{\textcolor{orange}{0.132}} \\
    AutoCas (GPT-2-Large) & 1.215 & 0.160 & 1.084 & 0.146 & 2.337 & 0.246 & 2.266 & 0.237 & 0.595 & 0.124 & 0.362 & 0.107 \\
    AutoCas (LLaMA-1B) & 1.203 & 0.154 & 0.864 & 0.132 & 2.248 & 0.231 & 2.235 & 0.229 & 0.502 & 0.098 & 0.259 & 0.064 \\
    AutoCas (LLaMA-3B) & 0.906 & 0.136 & 0.795 & 0.122 & 2.222 & 0.226 & 2.216 & 0.225 & 0.328 & 0.082 & 0.260 & 0.071 \\
    AutoCas (Qwen-3B) & 0.843 & 0.114 & 0.671 & 0.115 & 2.266 & 0.234 & 2.102 & 0.207 & 0.357 & 0.827 & 0.347 & 0.076 \\
    AutoCas (Qwen-7B) & \textbf{\textcolor[HTML]{cf2826}{0.776}} & \textbf{\textcolor[HTML]{cf2826}{0.112}} & \textbf{\textcolor[HTML]{cf2826}{0.624}} & \textbf{\textcolor[HTML]{cf2826}{0.109}} & \textbf{\textcolor[HTML]{cf2826}{2.190}} & \textbf{\textcolor[HTML]{cf2826}{0.203}} & \textbf{\textcolor[HTML]{cf2826}{1.973}} & \textbf{\textcolor[HTML]{cf2826}{0.182}} & \textbf{\textcolor[HTML]{cf2826}{0.290}} & \textbf{\textcolor[HTML]{cf2826}{0.076}} & \textbf{\textcolor[HTML]{cf2826}{0.138}} & \textbf{\textcolor[HTML]{cf2826}{0.055}} \\
    \midrule
    Improve & \textbf{66.38\%} & \textbf{44.28\%} & \textbf{71.35\%} & \textbf{49.07\%} & \textbf{65.65\%} & \textbf{46.01\%} & \textbf{56.26\%} & \textbf{41.48\%} & \textbf{73.48\%} & \textbf{63.29\%} & \textbf{86.35\%} & \textbf{75.98\%} \\
    \bottomrule
    \end{tabular}    
    \end{table*}

\subsection{Experimental settings}
    \noindent \textbf{Datasets.}
    To evaluate the performance of AutoCas, we employ two social networks: $\mathtt{Weibo}$~\cite{cao2017deephawkes} and $\mathtt{Twitter}$~\cite{weng2013virality} , and one citation network: $\mathtt{APS}$\footnote{\url{https://journals.aps.org/datasets}\label{footnote1}}.
    The target is to predict the number of retweets of posts in social networks and the citation count of academic papers.
    Statistics of the datasets are summarized in Table ~\ref{datasets}, and further details are provided in Appendix~\ref{datasetdetails}.

    \noindent \textbf{Setup and evaluation.}
    Following the previous studies~\cite{zhou2024modeling}, we randomly split each datasets into a training set (70\%), a validation set (15\%) and a test set (15\%).
    In the training phase, we use the training set for cascade modeling and prompt learning. 
    In the inference phase, we perform generation and popularity prediction on partially observed cascades.
    For evaluation metrics, we use Mean Square Log-transformed Error (MSLE) and Mean Absolute Percentage Error (MAPE) to evaluate the performance.
        
    \noindent \textbf{Baselines.}
    To evaluate AutoCas, we compare it with the state-of-the-art methods from three categories: Feature-based methods, Statistics-based methods and Deep learning-based methods. More baseline details can be found in Appendix~\ref{baselinedetails}.
    
    \noindent $\bullet$ Feature-based methods: Feat-P~\cite{szabo2010predicting}, Feat-Linear, and Feat-MLP.
    
    \noindent $\bullet$ Statistics-based methods: DeepHawkes~\cite{cao2017deephawkes}, TimeSeries~\cite{pinto2013using}.

    \noindent $\bullet$ Deep learning-based methods: 
    CasCN~\cite{CasCN}, CoupledGNN~\cite{cao2020popularity}, 
    FOREST~\cite{yang2021full},
    LatentODE~\cite{LatentODE},
    DyDiff-VAE~\cite{Dydiff-VAE},
    CTCP~\cite{CTCP}, MINDS~\cite{jiao2024enhancing}, CasFlow~\cite{casflow},  CasTFormer~\cite{Castformer},
    CasDO~\cite{CasDO}.

    \noindent \textbf{Implementation Details.} 
    We implement our model with the PyTorch framework on NVIDIA RTX 4090 GPU with 24GB of memory. We adopt several prevalent large language models (i.e., GPT-2, GPT-2-Large, LLaMA3.2-1B, LLaMA3.2-3B, Qwen2.5-3B, and Qwen2.5-7B) as our backbone. These pre-trained LLMs are downloaded from Huggingface\footnote{\url{https://huggingface.co/}}. 
    We set both of the local and global embedding dim  to 40.
    The maximum length $l$ of cascade tokens for the Weibo, Twitter, and APS datasets is set to 500, 400, and 200, respectively. The batch size is chosen from $\{64,128,256\}$.
    Adam is adopted for optimization and the patience in early stopping is set to 16.

    \subsection{Performance Comparison (RQ1)}
    We compare AutoCas against baselines across the three datasets.
    The main results are presented in Table \ref{results}. 
    From the results, we can observe that AutoCas significantly improves over state-of-the-art methods. 
    It achieves the lowest MSLE and MAPE values across all datasets, exhibiting a remarkable enhancement of approximately 50\% or even more compared to the current models.
    Notably, AutoCas is compatible with any decoder-only LLMs, demonstrating great generalization and applicability.
    Even in its worst outcomes, AutoCas consistently outperforms all existing methods by a substantial margin.
    In particular, on the APS dataset, AutoCas delivers an impressive 86.35\% improvement over CasDO on MSLE, underscoring its effectiveness.
    These results thoroughly verify the superiority of LLM-based predictor in information cascading modeling.


    \begin{figure}[htbp]
    \centering
    \begin{subfigure}[b]{\linewidth}
        \centering
        \includegraphics[width=\linewidth]{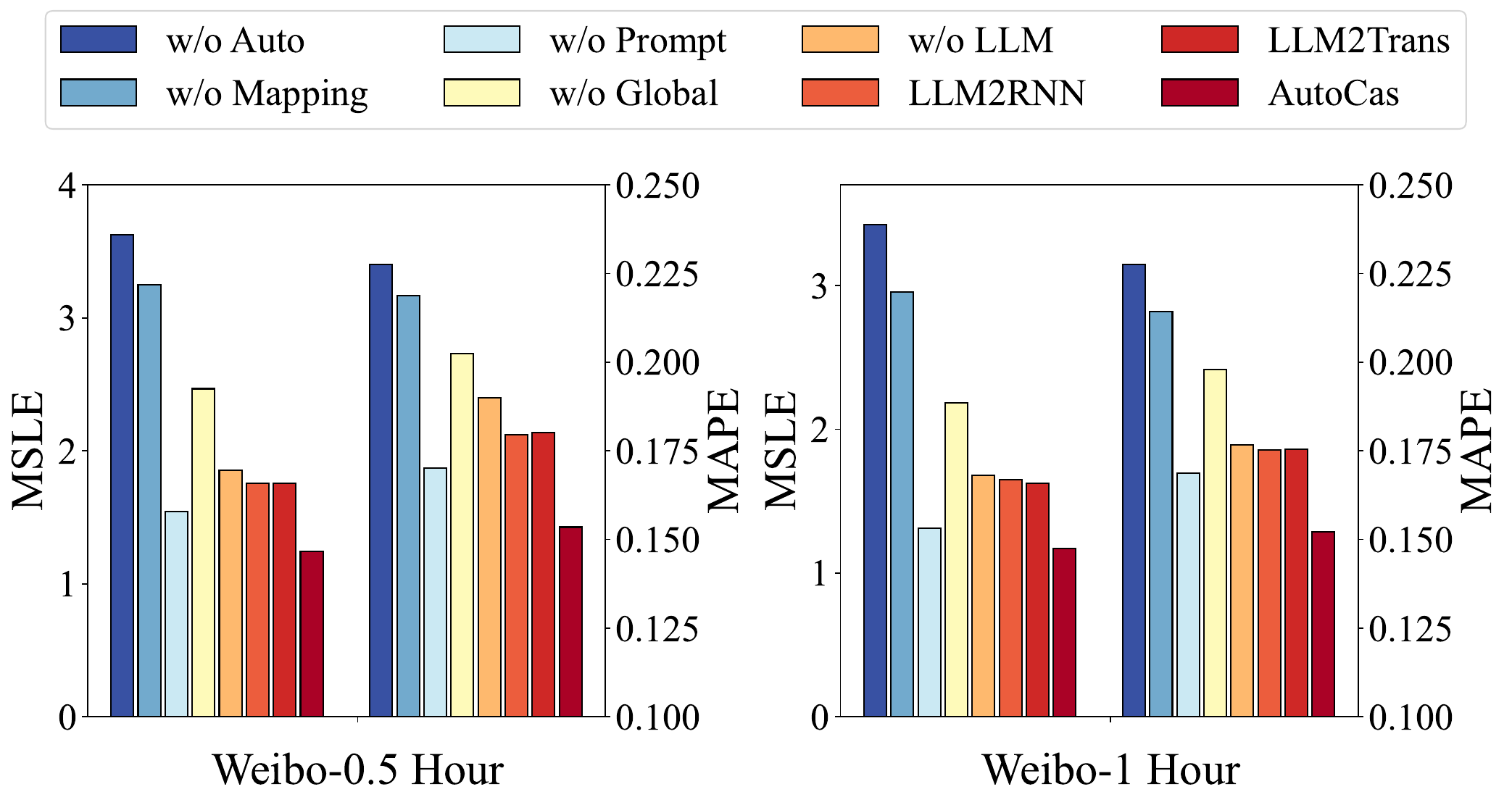}
        \caption{Ablation study results on Weibo dataset}
    \end{subfigure}

    \vskip\baselineskip 

    \begin{subfigure}[b]{\linewidth}
        \centering
        \includegraphics[width=\linewidth]{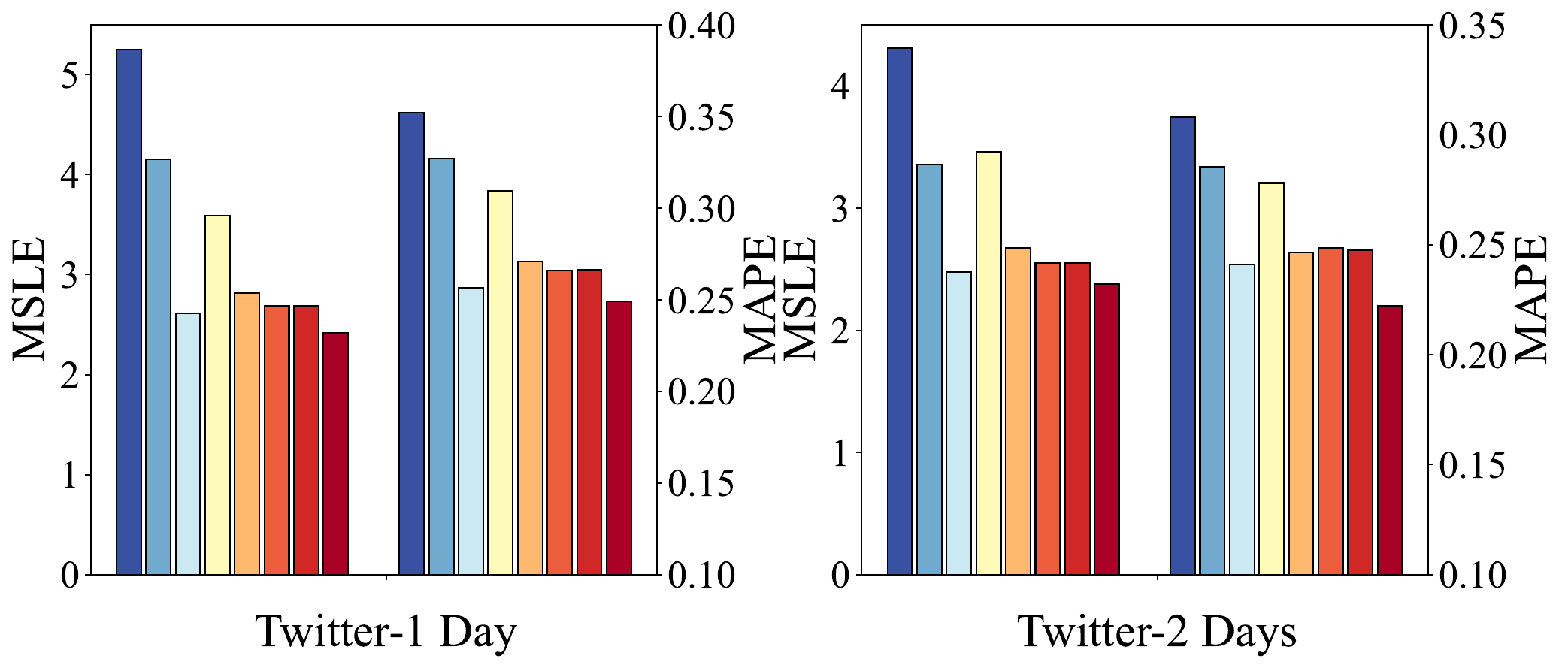}
        \caption{Ablation study results on Twitter dataset}
    \end{subfigure}

    \vskip\baselineskip 

    \begin{subfigure}[b]{\linewidth}
        \centering
        \includegraphics[width=\linewidth]{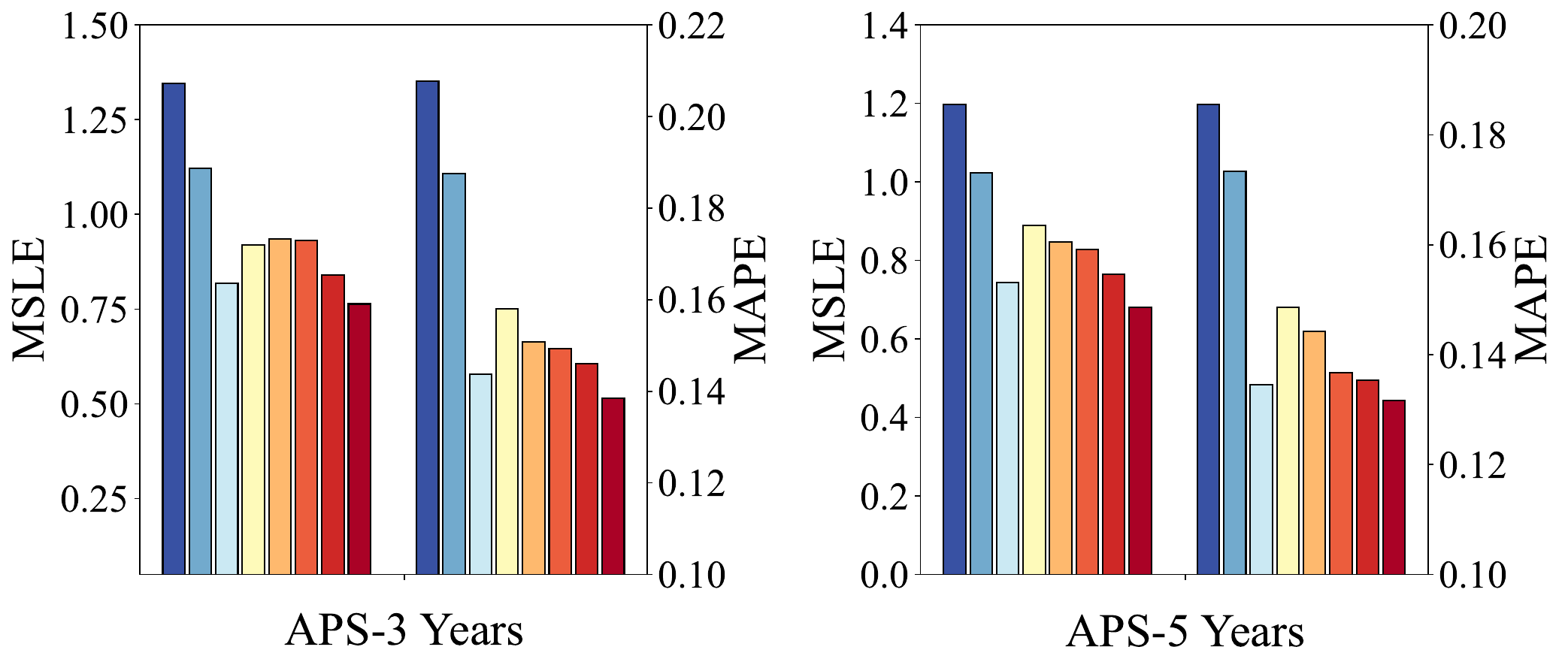}
        \caption{Ablation study results on APS dataset}
    \end{subfigure}
    
    \caption{Ablation study results of AutoCas and its variants on three datasets.}
    \label{ablation}
    \end{figure}

    \subsection{Ablation Study (RQ2)}
    To validate the effectiveness of the individual components in our model, we compared the full model with GPT-2 as backbone against seven variants: 
    (1) \textit{w/o Auto} employs a non-autoregressive approach which receives only the cascade sequence within observation time as input and directly outputs the popularity at prediction time; 
    (2) \textit{w/o Prompt} removes prompt; 
    (3) \textit{w/o Mapping} removes the MLP projector and adapter layer, only using single linear layer; 
    (4) \textit{w/o Global} removes the global embeddings in cascade tokenization, only using local embeddings; 
    (5) \textit{w/o LLM} removes the large language model and passing input tokens directly to the last layer; 
    (6) \textit{LLM2Trans} replaces the large language model with a single transformer block; 
    (7) \textit{LLM2RNN} replaces the large language model with a single RNN block.
    As shown in Figure \ref{ablation}, we can make the following observations.

    \noindent $\bullet$ \textbf{Effectiveness of global embedding.} 
    The performance of \textit{w/o Global} reveals the importance of including global embedding to provide a more comprehensive user embedding, facilitating subsequent cascade token sequences construction.
    
    \noindent $\bullet$ \textbf{Effectiveness of components in autoregressive cascade modeling.} 
    The results of \textit{w/o Auto} are clearly worse than those of AutoCas and all other variants, which indicates that even without leveraging the powerful capabilities of LLMs, solely relying on the autoregressive method for cascade prediction yields substantially better performance than approaches that forgo autoregression.
    The poor performance of \textit{w/o Mapping} suggests that the linear layer is overly simplistic, leading to limited representational capacity insufficient for mapping cascades into embeddings interpretable by LLMs. 
    In contrast, the MLP, with its multiple layers of non-linear transformations, captures complex patterns more effectively, enhancing the LLM’s expressiveness and predictive performance.
    
    \noindent $\bullet$ \textbf{Effectiveness of LLMs as backbone.}
    The performance of \textit{w/o LLM} is notably inferior to that of \textit{AutoCas}, which fully reveals the importance of large language model.
    The results of \textit{LLM2RNN} and \textit{LLM2Trans} are also fall short compared to that of \textit{AutoCas}, despite the fact that the LLM parameters in \textit{AutoCas} are fully frozen, while the backbones of \textit{LLM2RNN} and \textit{LLM2Trans} are trainable network structures. 
    This highlights that replacing the LLM with an alternative architecture, even with the addition of learnable parameters, results in a significant performance decline, thereby thoroughly demonstrating the effectiveness of LLMs.
    
    \noindent $\bullet$ \textbf{Effectiveness of prompt learning.}
    The performance of \textit{w/o Prompt} reveals the improvement achieved by the skillful textual prompt design.
    This validates its effectiveness in facilitating LLM to make precise predictions.


    \begin{figure*}[htbp]
    \centering
    \subfloat[MSLE loss on Weibo dataset]{
    \includegraphics[width=0.32\linewidth]{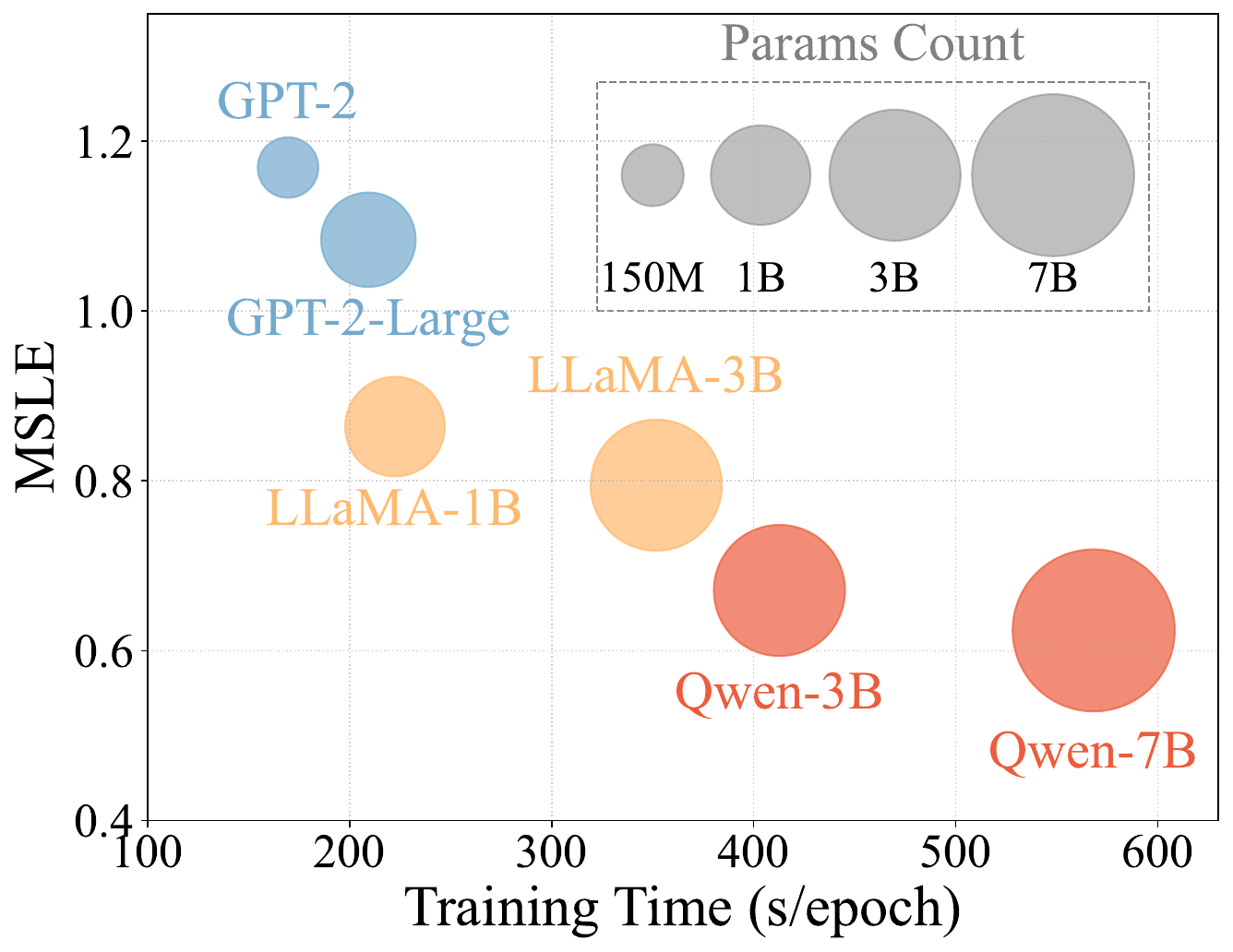}
    }
    \subfloat[MSLE loss on Twitter dataset]{
    \includegraphics[width=0.32\linewidth]{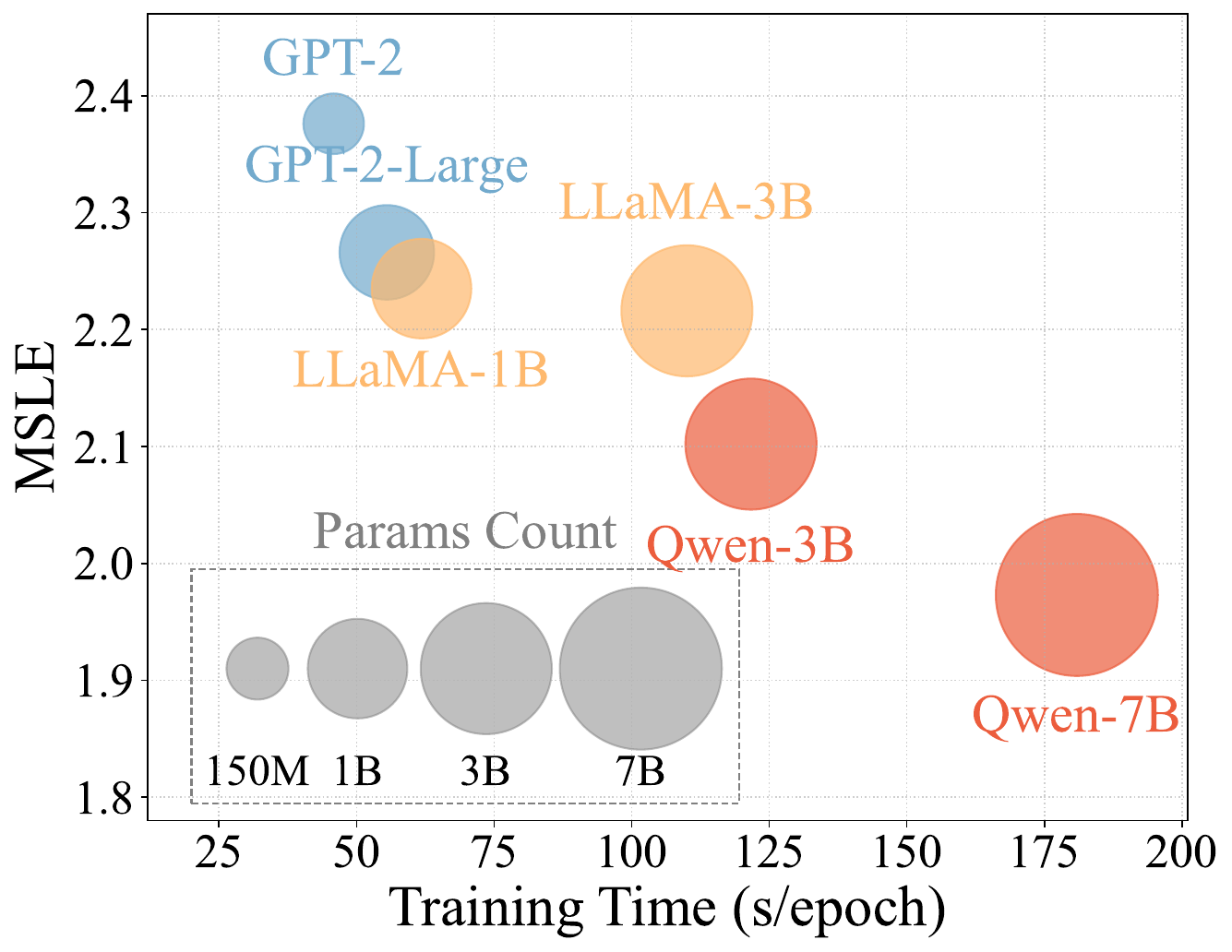}
    }
    \subfloat[MSLE loss on APS dataset]{
    \includegraphics[width=0.32\linewidth]{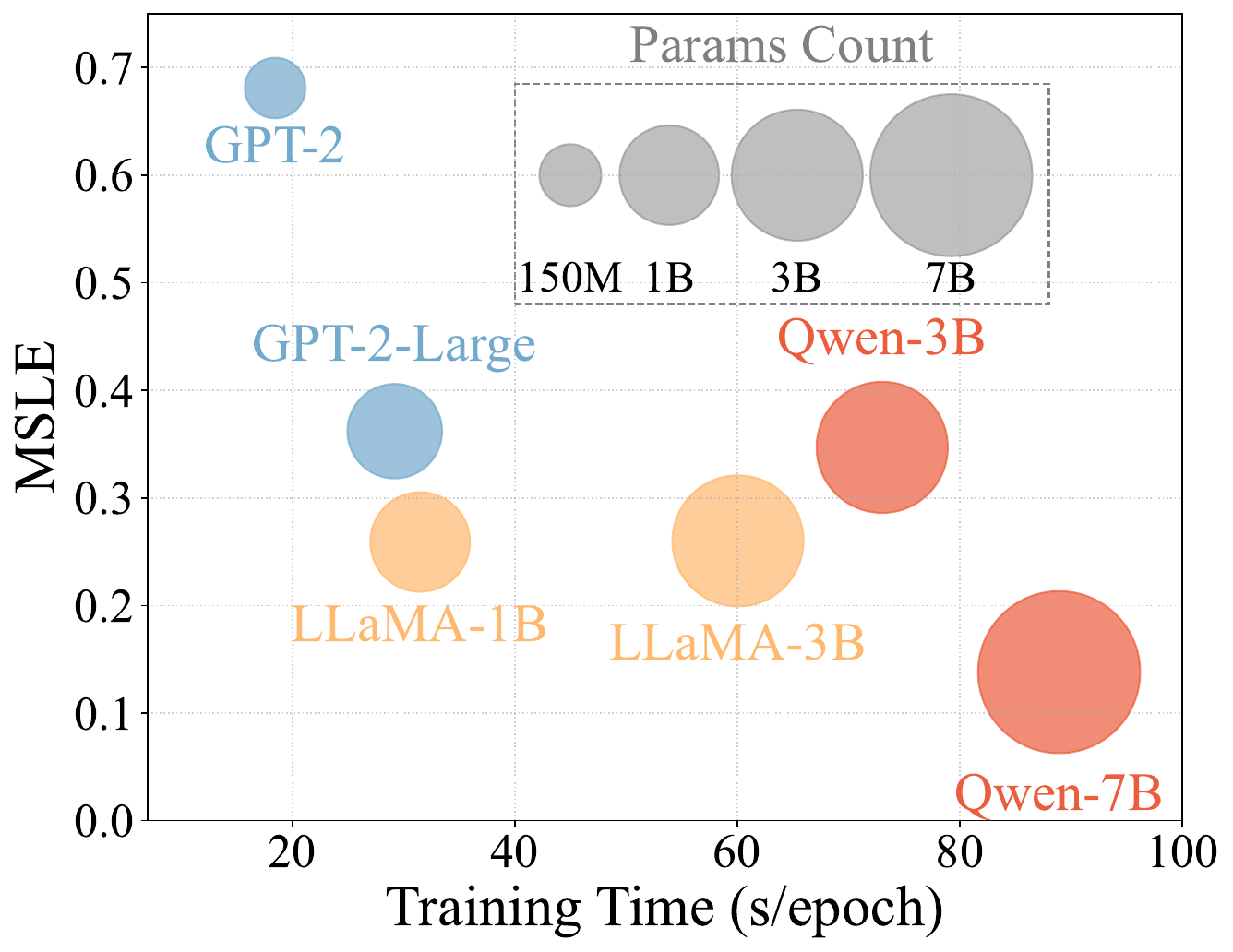}
    }
    \caption{Performance and training efficiency comparison of AutoCas with different LLMs across three datasets (Weibo: 1 Hour, Twitter: 2 Days, and APS: 5 Years), evaluated under the same batch size and learning rate.}
    \label{scaling}
    \end{figure*}

    \subsection{Scaling Behavior (RQ3)}
    Scalability is an crucial characteristic that evolves from small models to large foundation models. 
    In this section, we explore the scaling trends of AutoCas in cascade popularity prediction.
    From the results in Table \ref{results}, we observe that the prediction accuracy generally improves with the increase in LLM parameters, and the largest Qwen-7B consistently delivers optimal predicting performance. 
    These findings validate the scaling behavior of AutoCas empowered by LLMs in cascade popularity prediction.
   
    This scaling behavior of LLM-based predictors introduces a trade-off between performance and computational efficiency. To provide a comprehensive assessment, we evaluate each adapted predictor from three perspectives: performance, training speed, and parameters, using the same batch size and learning rate, as illustrated in Figure \ref{scaling}. Additional experiments can be found in Appendix \ref{appendixD}.
    One can observe that the models generally achieve lower MSLE as the number of parameters increases. However, this improvement comes at the cost of longer training times, as seen with larger models like Qwen-7B. While smaller models such as GPT-2 train faster but exhibit higher MSLE, indicating limited effectiveness.
    Therefore, selecting the optimal model requires balancing accuracy gains against computational efficiency to ensure practical deployment within resource constraints.

    \subsection{Efficiency Analysis (RQ4)}
\begin{figure}[htbp]
    \centering
    \subfloat[Convergence speed on Weibo]{
    \label{convergence}
    \includegraphics[width=0.49\linewidth]{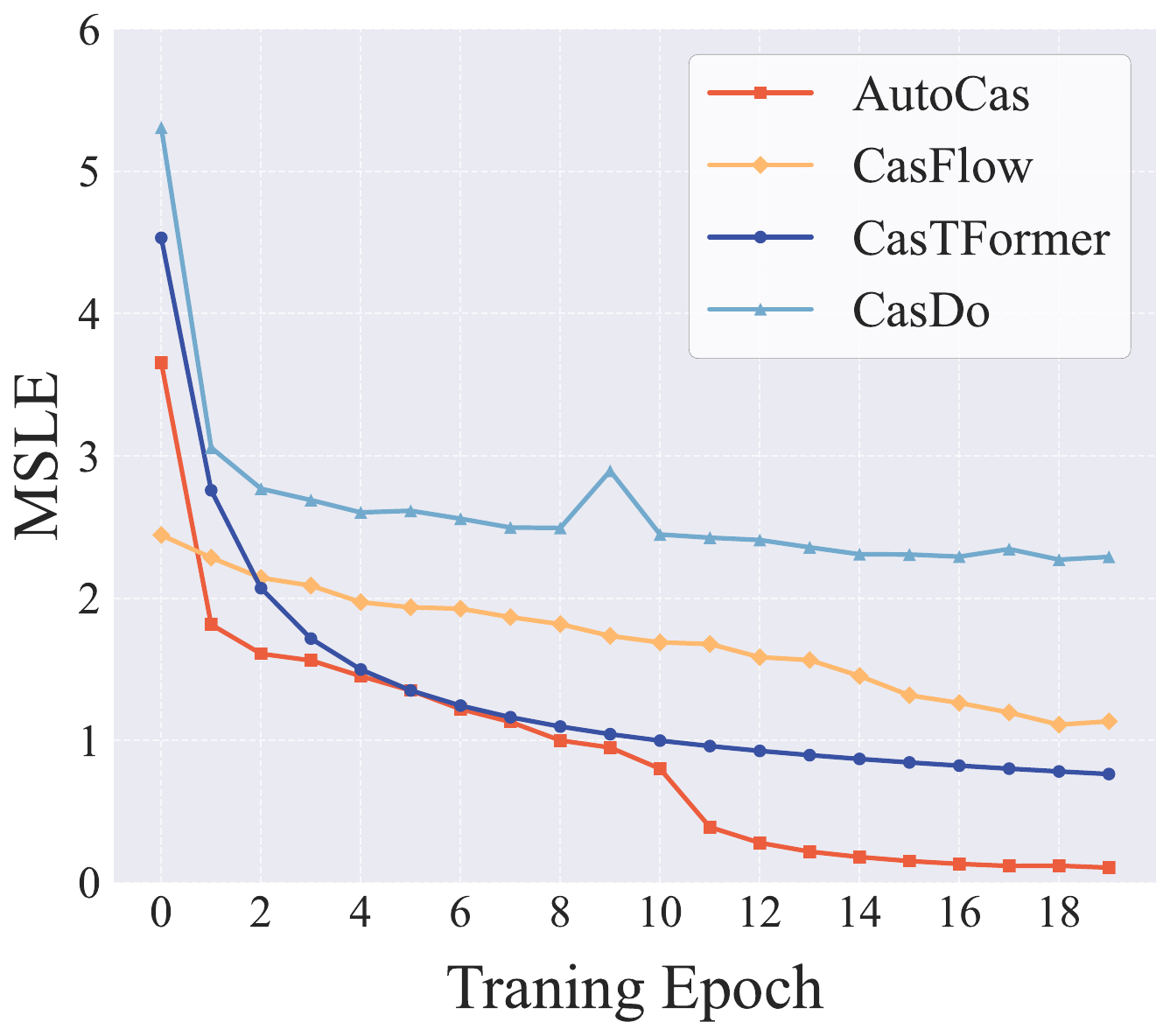}
    }
    \subfloat[Inference time on APS]{
    \label{inference}
    \includegraphics[width=0.49\linewidth]{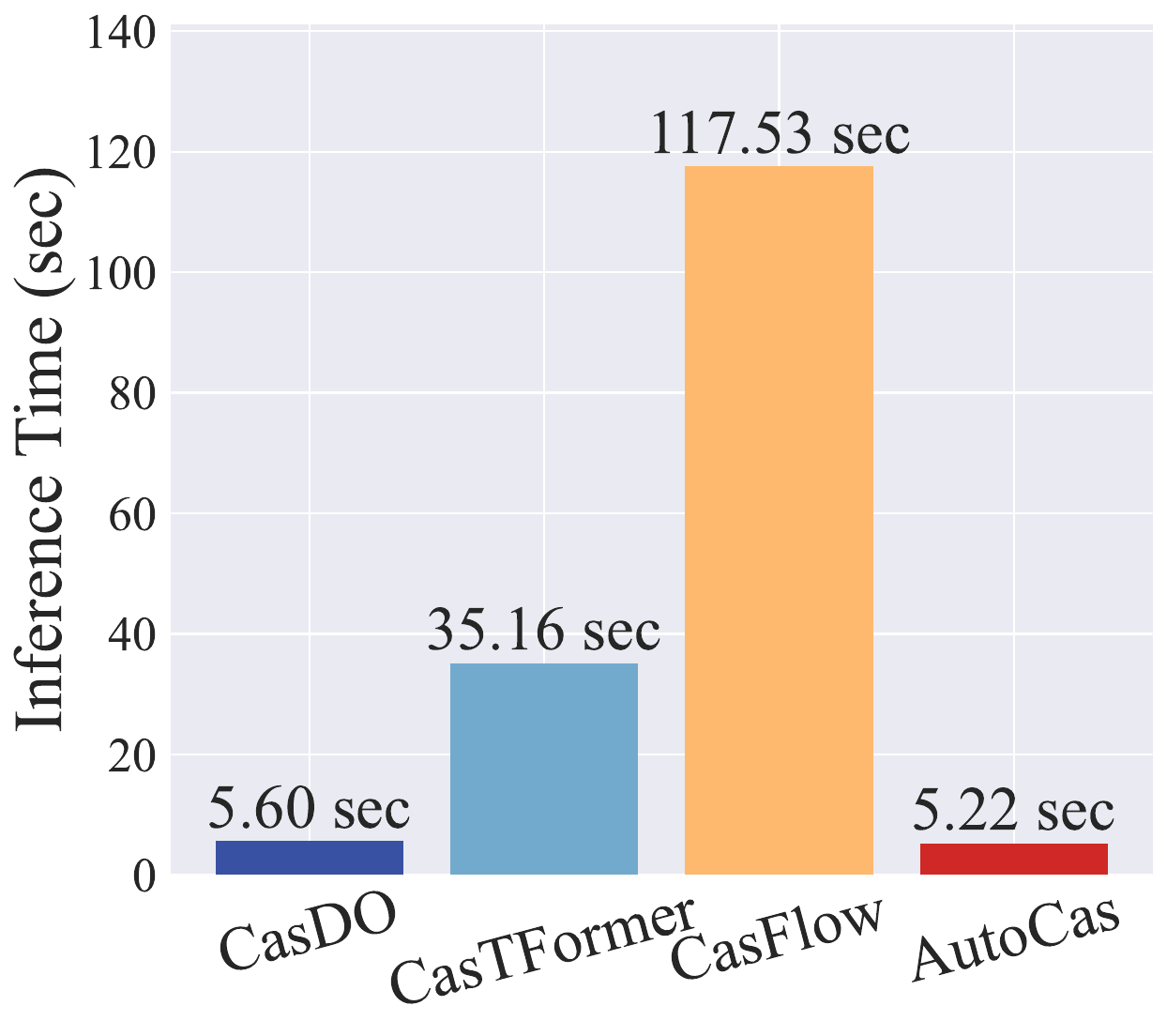}
    }
    \caption{Convergence speed and inference time of AutoCas with GPT-2 as backbone and baselines on Weibo and APS. }
    \label{RQ4_1}
    \end{figure}

\begin{figure}[htpb]
    \includegraphics[width=0.99\linewidth]{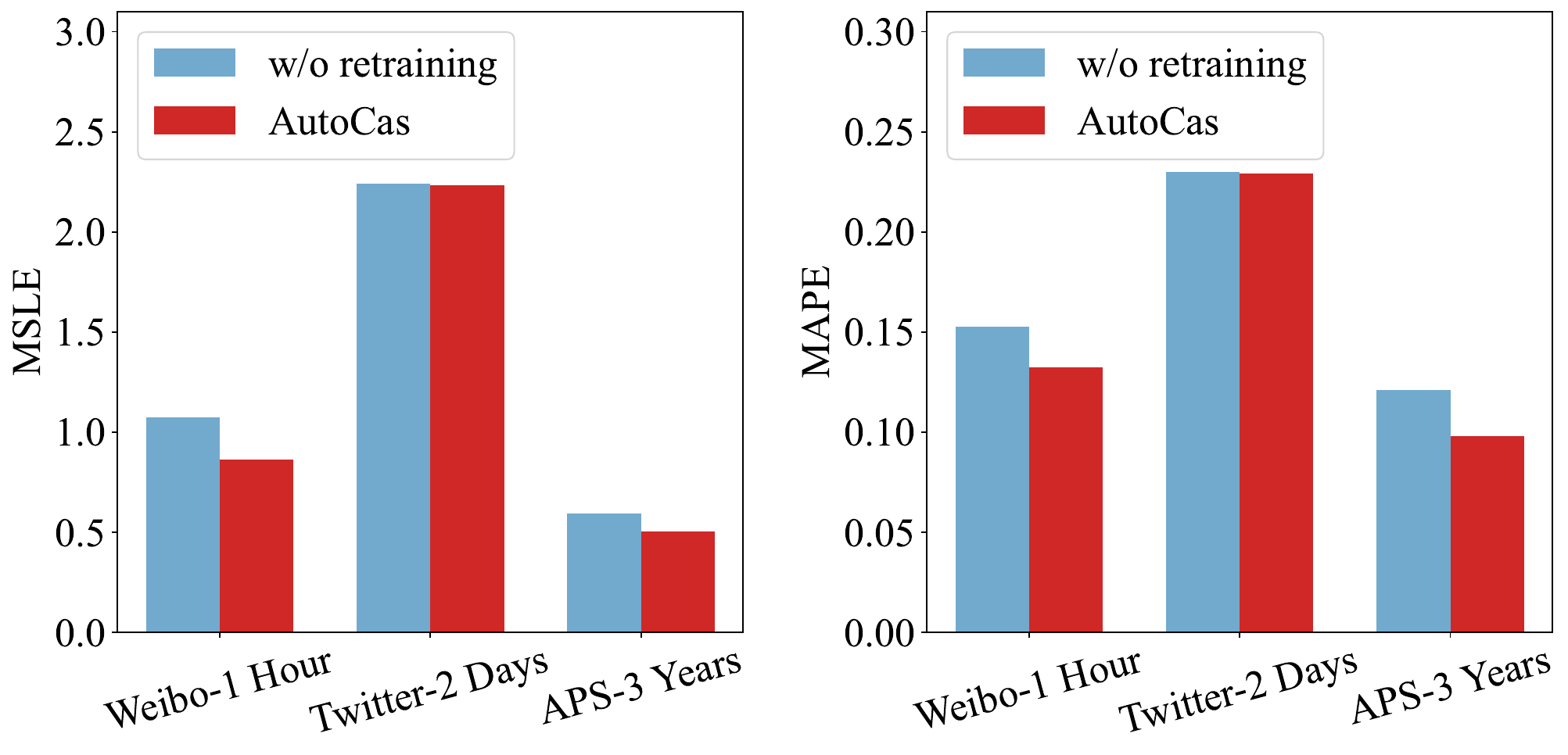}
    \caption{Performance comparison between \textit{AutoCas} and \textit{w/o retraining} with LLaMA-1B as backbone. 
    \textit{AutoCas} denotes that training and inference are on the same time partition.
    \textit{w/o retraining} indicates that the model is trained on a finer-grained time partition (Weibo: 0.5 hour, Twitter: 1 day, APS: 1 year) and directly applied on a coarser-grained time partition (Weibo: 1 hour, Twitter: 2 days, APS: 3 years).}
    \label{woretraining}
    \end{figure}

    To evaluate the time efficiency of AutoCas, especially in comparison to SOTA baselines, we record the training convergence speed and inference time for AutoCas and each baseline on the Weibo and APS dataset, with the results shown in Figure \ref{RQ4_1}.
    Then we discuss the cross time partition efficiency of AutoCas, and the results are presented in Figure \ref{woretraining}.
    Additionally, we examine the paramter efficiency of AutoCas, which is depicted in Table \ref{RQ4_2}.
    Based on these results, we can draw the following conclusions:
    
    \noindent $\bullet$ \textbf{AutoCas converges faster than other methods with less number of training epochs.}
    Notably, compared to other models, AutoCas exhibits a rapid decline in training loss, reaching a low value and converging within approximately 12 epochs.

    \noindent $\bullet$ \textbf{AutoCas inferences faster than other methods.}
    Despite adopting the autoregressive paradigm and the backbone with a large-scale LLM, AutoCas still maintains rapid inference speed, outperforming even models with simpler architectures in efficiency.

    \begin{table}[htbp]
    \centering
    \caption{Parameter statistics of Autocas with different LLMs on Weibo and APS datasets, including the ratio of learnable parameters relative to the total parameters.}
    \small
    \label{RQ4_2}
    \begin{tabular}{c | c | c | c | c}
        \toprule
        Datasets & LLMs & Learnable (M)  & Total (M) & Ratio (\%)     \\
        \midrule
        \multirow{6}{*}{Weibo} 
                  & GPT-2   & 31.155  & 155.594  & 20.023\% \\
                  & GPT-2-Large   & 31.672  & 1094.670  & 2.893\% \\
                  & LLaMA-1B & 31.811  & 1267.626 & 2.510\% \\
                  & LLaMA-3B & 32.337  & 3245.086 & 0.997\% \\
                  & Qwen-3B  & 31.811  & 3117.750 & 1.020\% \\
                  & Qwen-7B  & 32.599  & 7648.216 & 0.426\% \\
        \midrule
        \multirow{6}{*}{APS} 
                  & GPT-2   & 12.699  & 137.138  & 9.260\% \\
                  & GPT-2-Large   & 13.187  & 1078.699  & 1.222\% \\
                  & LLaMA-1B & 13.355  & 1249.170 & 1.069\% \\
                  & LLaMA-3B & 13.880  & 3227.630 & 0.430\% \\
                  & Qwen-3B  & 13.355  & 3099.294 & 0.431\% \\
                  & Qwen-7B  & 14.143  & 7630.760 & 0.185\% \\
        \bottomrule
    \end{tabular}
    \end{table}

     \noindent $\bullet$ \textbf{AutoCas can be trained once at a fine-grained time partition and directly applied to coarser time divisions for prediction.}
    Unlike existing methods that necessitate retraining for each distinct observation time, AutoCas only needs to be trained once on a small time division. By adaptively modifying the input length during inference, it can efficiently predict cascade population at any observed time.
    Notably, the performance achieved without retraining is highly comparable to that of direct training, demonstrating the impressive capability and efficiency of the autoregressive training strategy employed by AutoCas.
    
    \noindent $\bullet$ \textbf{The learnable parameters of AutoCas constitute only a small fraction of the total.}
    As the number of LLM parameters increases, the total parameters in Autocas rise significantly, while the trainable parameters remain consistently low.
    Consequently, as the model scales, AutoCas not only achieves substantial improvements in performance but also exhibits a decreasing ratio of trainable parameters relative to the total, underscoring its efficiency.
    
\section{Conclusion}
    In this paper, we present AutoCas, a novel framework that repurposes LLMs as predictors of information cascade.
    We tokenize the cascade data and align the cascade diffusion process with language generation at the token level.
    Then we project cascade tokens into the embedding space of LLMs, and propose the paradigm of autoregressive cascade modeling to fully leverage the potential of LLM architecture.
    Further, we adopt prompt learning in cascade popularity prediction, and design prompt tokens to facilitate specific task adaptation.
    Finally, our experimental results demonstrate the superior performance of AutoCas in the cascade popularity prediction task, and it exhibits the scaling behavior empowered by LLMs.
    In future work, we will explore the unified paradigm for integrating social data and natural language, and the possibility of building foundation models in social computing.


\bibliographystyle{unsrt}
\bibliography{ref}

\appendix

\section{Further Details of Datasets}
    \label{datasetdetails}
    \noindent $\textbf{Weibo.}$ 
    This cascade dataset is released in~\cite{cao2017deephawkes}, which contains Chinese tweets and retweets of Sina Weibo on 1 June, 2016.
    The global graph is built  by all user retweeting relationships from Sina Weibo.
    The observation time $t_o$ is set as 0.5 hour or 1 hour, and the prediction time $t_p$ is set as 24 hours.
    
    \noindent $\textbf{Twitter.}$ 
    This hashtag cascade dataset is released in~\cite{weng2013virality}, which contains English tweets on the social platform from 24 March to 25 April in 2012. 
    A hashtag and its participants form an information cascade.
    The global graph is built by hashtag participants and their interactions, including follower, followee, retweet and mention.
    The observation time $t_o$ is set as 1 day or 2 days, and the prediction time $t_p$ is set as 32 days.

    \noindent $\textbf{APS.}$ 
    This citation cascade dataset is released by American Physical Society\textsuperscript{\ref{footnote1}}, which contains 616,316 scientific papers published by APS.
    Each Paper and its citations form a citation cascade.
    The global graph is built by the author interactions.
    The observation time $t_o$ is set as 3 years or 5 years, and the prediction time $t_p$ is set as 20 years. 


    \begin{figure*}[htbp]
    \centering
    \subfloat[MAPE loss on Weibo dataset]{
        \includegraphics[width=0.33\linewidth]{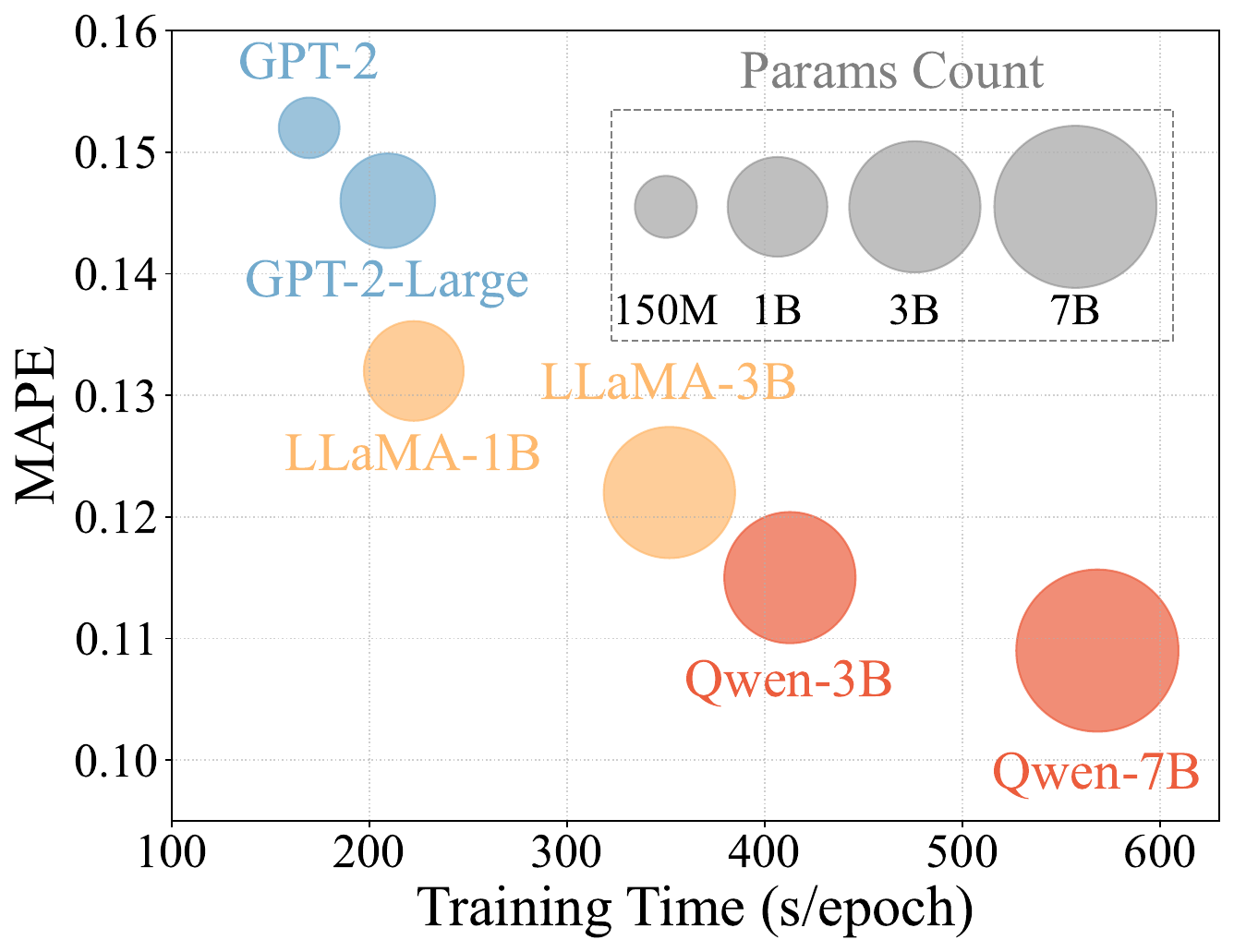}
    }
    \subfloat[MAPE loss on Twitter dataset]{
        \includegraphics[width=0.33\linewidth]{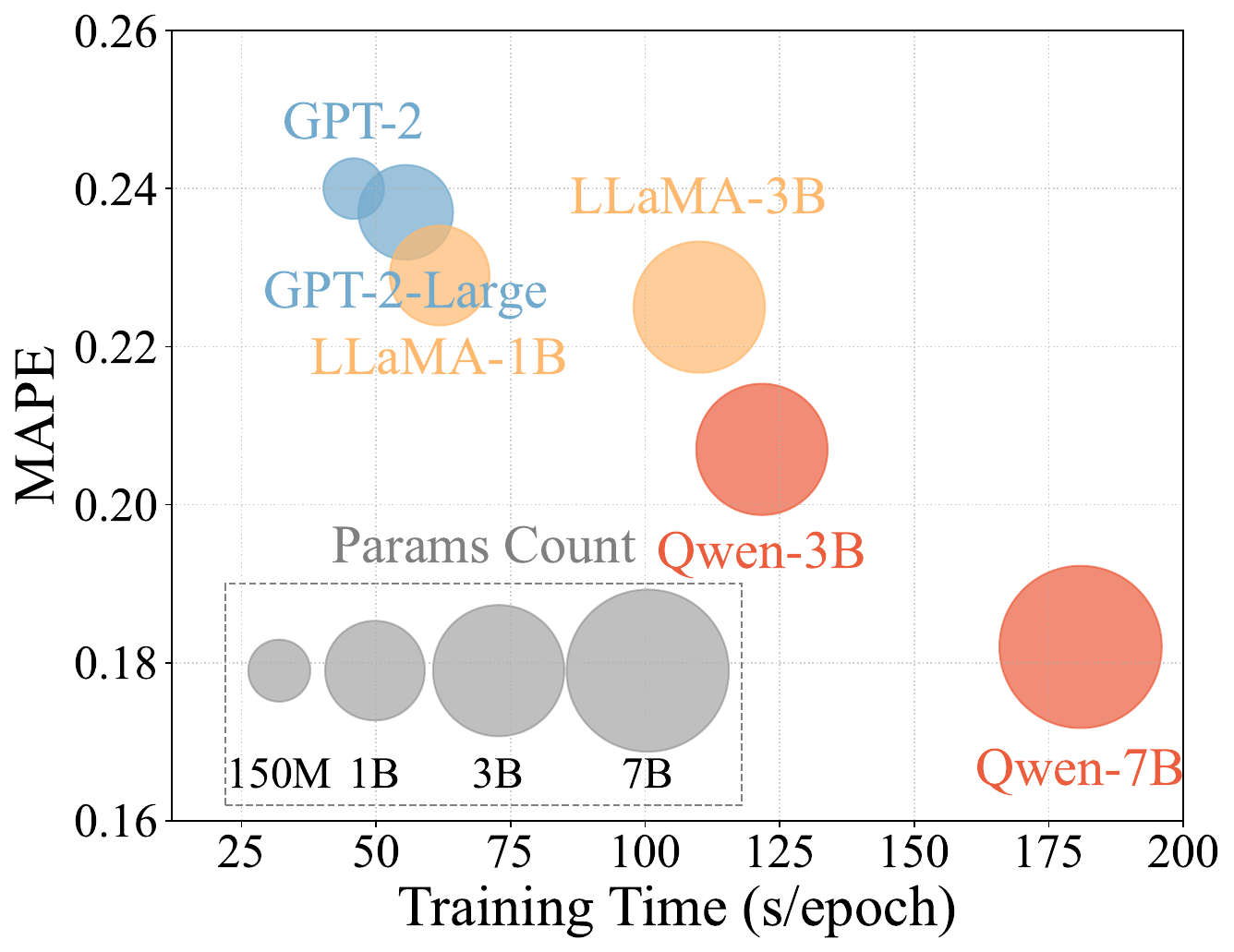}
    }
    \subfloat[MAPE loss on APS dataset]{
        \includegraphics[width=0.33\linewidth]{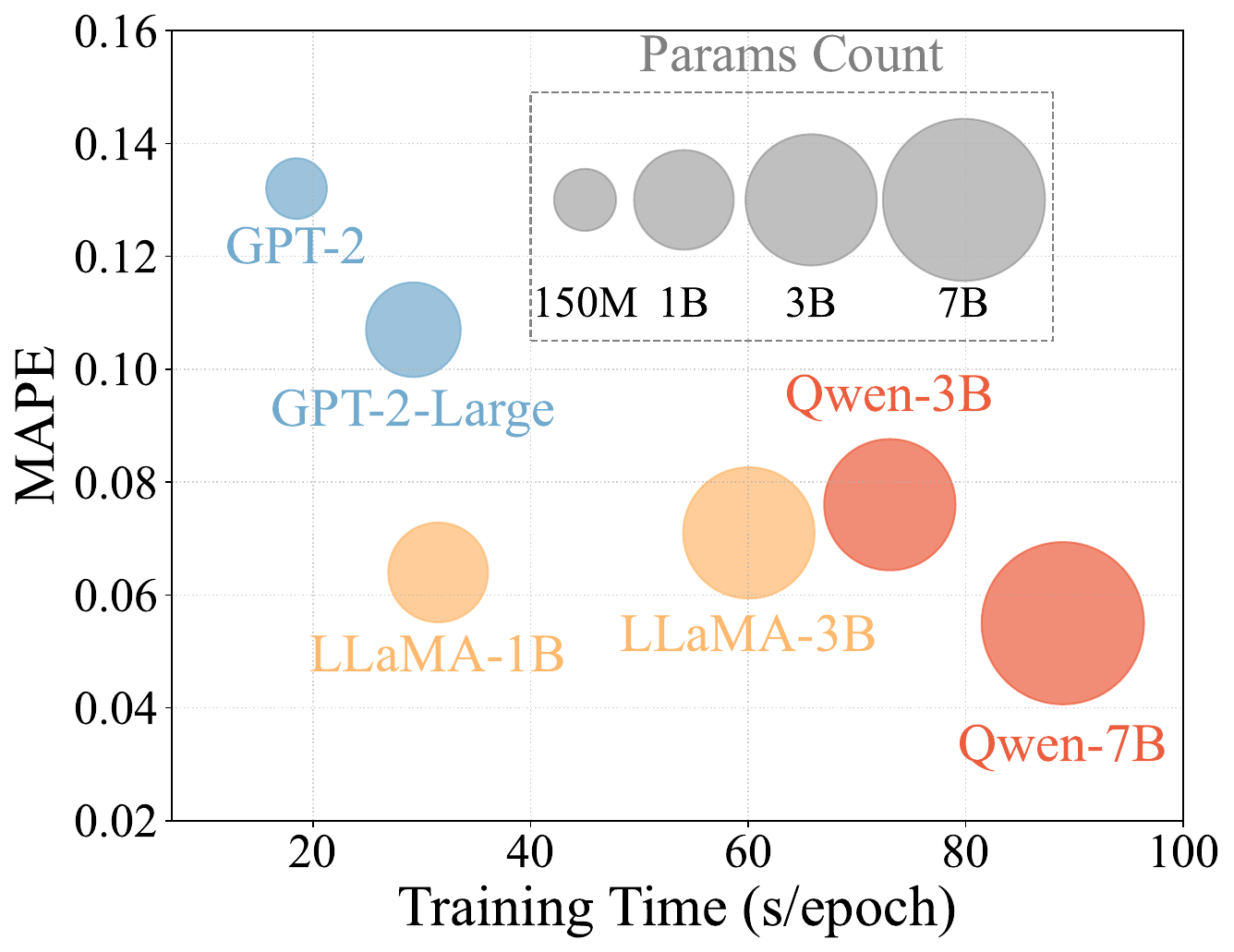}
    }
    \caption{Performance on MAPE and training efficiency comparison of AutoCas with different LLMs across three datasets(Weibo: 1 Hour, Twitter: 2 Days, and APS: 5 Years), evaluated under the same batch size and learning rate.}
    \label{scaling2}
    \end{figure*}
    
\section{Further Details of Baselines}
    \label{baselinedetails}
     \noindent \textbf{Baselines.}
     To assess the effectiveness and efficiency of AutoCas, we conducted a comparative analysis with 15 leading-edge models in the domain of cascade popularity prediction. 
     The models we benchmarked against are as follows: 
     
    
    
 
    \noindent $\bullet$ \textbf{Feature-P}~\cite{szabo2010predicting} uses the observed popularity to predict future popularity, and models the dynamics using machine learning methods.
    
    \noindent $\bullet$ \textbf{Feature-Linear} follows the method used in ~\cite{CasCN} for extracting structural and temporal cascade features, and uses a linear regression model to predict cascade popularity.
    
    \noindent $\bullet$ \textbf{Feature-MLP} is a variant of Feature-Linear, and uses a two-layer MLP model to predict cascade popularity.
    
    \noindent $\bullet$ \textbf{DeepHawkes}~\cite{cao2017deephawkes}uses deep learning techniques to simulate the factors of Hawkes process for popularity prediction, including the influence of users, self-exciting mechanism, and time decay.

    \noindent $\bullet$ \textbf{TimeSeries}~\cite{pinto2013using} 
    builds time series models to make cascade popularity prediction.
    
    \noindent $\bullet$ \textbf{CasCN}~\cite{CasCN} transforms the cascade graph into a sequence of cascade graph episodes, uses a Graph Convolution Network (GCN) on each episode to capture local structure, and then employs LSTM to model the dynamic of the sequence.
    
    \noindent $\bullet$ \textbf{CoupledGNN}~\cite{cao2020popularity} uses two coupled GNNs to model the interactions between users and the spread influence for cascade popularity prediction.
    
    \noindent $\bullet$ \textbf{FOREST}~\cite{yang2021full} combines reinforcement learning with a GRU-based micro-cascade sequence prediction model to solve the multi-scale cascade prediction problem.

    \noindent $\bullet$ \textbf{LatentODE}~\cite{LatentODE} generalizes discrete RNNs to continuoustime hidden dynamics defined by ODEs. It considers the latent representation a time-series variable in RNNs, and therefore is capable of handling arbitrary time gaps between observations
    
    \noindent $\bullet$ \textbf{DyDiff-VAE}~\cite{Dydiff-VAE} learns user interest evolution using GRU and estimates the propagation likelihood with a dual attentive variational autoencoders.

    \noindent $\bullet$ \textbf{CTCP}~\cite{CTCP} designs a graph learning framework for cascade popularity prediction by modeling cascades through a universal sequence of user-cascade and user-user interactions, and subsequently extracting the dynamic states of users and cascades from graph sequences.
    
    \noindent $\bullet$ \textbf{MINDS}~\cite{jiao2024enhancing} constructs sequential hypergraphs based on dynamic cascade interactions and integrates adversarial training to mitigate feature redundancy for enhanced popularity prediction..
    
    \noindent $\bullet$ \textbf{CasFlow}~\cite{casflow} learns cascade structural information from both local and global views, and uses Variational Auto-Encoder (VAE) and normalizing flows to enhance the learned cascade representations.
    
    \noindent $\bullet$ \textbf{CasTformer}~\cite{Castformer} incorporates a global spatio-temporal positional encoding and relative relationship bias matrices into the self-attention architecture, enabling the model to extract diverse cascade relationships for enhanced popularity prediction.
     
    \noindent $\bullet$ \textbf{CasDO}~\cite{CasDO} devises a temporal ODE network to extend discrete state transitions in RNNs to continuous-time dynamics, and introduces a probabilistic diffusion model to account for uncertainties in information diffusion.
    
\section{Further Details of Textual Prompt Template}
    \label{appendixC}
    In this section, we present the textual prompt template utilized in this study as follow:
    
    \begin{tcolorbox}[title=Textual Prompt Template for Weibo, colframe=black!30, colback=gray!5, left=0pt]
    \begin{lstlisting}[basicstyle=\ttfamily\small, xleftmargin=0pt, framexleftmargin=0pt]
    <|Start_prompt|>
    Dataset description:
    This is the cascade dataset of Weibo, 
    which contains Chinese tweets and retweets 
    of Sina Weibo. Its users and their reteets 
    behaviors form an information cascade.
    
    Data Description:
    This is the n-th information cascade token.
    
    Task Description:
    Forcast the next token given the previous
    token information, and we want to predict 
    the cascade size at final time.
    <|End_prompt|> 
    \end{lstlisting}
    \end{tcolorbox}

    \begin{tcolorbox}[title=Textual Prompt Template for Twitter, colframe=black!30, colback=gray!5, left=0pt]
    \begin{lstlisting}[basicstyle=\ttfamily\small, xleftmargin=0pt, framexleftmargin=0pt]
    <|Start_prompt|>
    Dataset description:
    This is the cascade dataset of Twitter, 
    which contains English tweets and retweets 
    of Twitter. A hashtag and its participants 
    form an information cascade.

    Data Description:
    This is the n-th information cascade token.
    
    Task Description:
    Forcast the next token given the previous
    token information, and we want to predict 
    the cascade size at final time.
    <|End_prompt|> 
    \end{lstlisting}
    \end{tcolorbox}

    \begin{tcolorbox}[title=Textual Prompt Template for APS, colframe=black!30, colback=gray!5, left=0pt]
    \begin{lstlisting}[basicstyle=\ttfamily\small, xleftmargin=0pt, framexleftmargin=0pt]
    <|Start_prompt|>
    Dataset description:
    This is the cascade dataset of APS, which 
    contains the citation cascade built by 
    author intheractions.

    Data Description:
    This is the n-th information cascade token.
    
    Task Description:
    Forcast the next token given the previous
    token information, and we want to predict 
    the cascade size at final time.
    <|End_prompt|> 
    \end{lstlisting}
    \end{tcolorbox}
    \noindent where \enquote{$n$} in textual prompt denotes the index of tokens.
    Note that the design of textual prompt is flexible, here we just provide a basic template, and leave more advanced designs for future work.

\section{Further Details of Additional Experiments}
\label{appendixD}
    This scaling behavior of LLM-based predictors introduces a trade-off between performance and computational efficiency. To provide a comprehensive assessment, we evaluate each adapted predictor from three perspectives: performance on MSLE, training speed, and parameter efficiency, as shown in Figure \ref{scaling}.

    In addition to MSLE, we also conduct the same experiments using MAPE, with the results presented in Figure \ref{scaling2}. The observed scaling behavior remains consistent with the findings based on MSLE.





    

\end{document}